\documentclass[%
 aip,
 amsmath,amssymb,
 reprint,%
]{revtex4-1}

\usepackage{graphicx}
\usepackage{dcolumn}
\usepackage{bm}

\usepackage[utf8]{inputenc}
\usepackage[T1]{fontenc}
\usepackage{mathptmx}
\usepackage{etoolbox}
\usepackage{bigints}
\usepackage[dvipsnames]{xcolor}
\usepackage{comment} 
\usepackage{multirow}

\makeatletter
\def\@email#1#2{%
 \endgroup
 \patchcmd{\titleblock@produce}
  {\frontmatter@RRAPformat}
  {\frontmatter@RRAPformat{\produce@RRAP{*#1\href{mailto:#2}{#2}}}\frontmatter@RRAPformat}
  {}{}
}%
\makeatother
\begin{document}

\preprint{AIP/123-QED}

\title[Homogeneous nucleation of carbon dioxide hydrate]{Homogeneous nucleation rate of carbon dioxide hydrate formation under experimental condition from Seeding simulations}

\author{I. M. Zer\'on}
\affiliation{Laboratorio de Simulaci\'on Molecular y Qu\'imica Computacional, CIQSO-Centro de Investigaci\'on en Qu\'imica Sostenible and Departamento de Ciencias Integradas, Universidad de Huelva, 21006 Huelva Spain}

\author{J. Algaba}
\affiliation{Laboratorio de Simulaci\'on Molecular y Qu\'imica Computacional, CIQSO-Centro de Investigaci\'on en Qu\'imica Sostenible and Departamento de Ciencias Integradas, Universidad de Huelva, 21006 Huelva Spain}

\author{J. M. M\'{\i}guez}
\affiliation{Laboratorio de Simulaci\'on Molecular y Qu\'imica Computacional, CIQSO-Centro de Investigaci\'on en Qu\'imica Sostenible and Departamento de Ciencias Integradas, Universidad de Huelva, 21006 Huelva Spain}

\author{J. Grabowska}
\affiliation{Department of Physical Chemistry, Faculty of Chemistry, Gdansk University of Technology, ul. Narutowicza 11/12, 80-233 Gdansk, Poland}

\author{S. Blazquez}
\affiliation{Dpto. Qu\'{\i}mica F\'{\i}sica, Fac. Ciencias Qu\'{\i}micas, Universidad Complutense de Madrid, 28040 Madrid, Spain}

\author{E. Sanz}
\affiliation{Dpto. Qu\'{\i}mica F\'{\i}sica, Fac. Ciencias Qu\'{\i}micas, Universidad Complutense de Madrid, 28040 Madrid, Spain}

\author{C. Vega}
\affiliation{Dpto. Qu\'{\i}mica F\'{\i}sica, Fac. Ciencias Qu\'{\i}micas, Universidad Complutense de Madrid, 28040 Madrid, Spain}

\author{F. J. Blas$^{*}$}
\affiliation{Laboratorio de Simulaci\'on Molecular y Qu\'imica Computacional, CIQSO-Centro de Investigaci\'on en Qu\'imica Sostenible and Departamento de Ciencias Integradas, Universidad de Huelva, 21006 Huelva Spain}

\begin{abstract}

We investigate the nucleation of carbon dioxide (CO$_2$) hydrates from carbon dioxide aqueous solutions by means of molecular dynamics simulations
using the TIP4P/Ice and the TraPPE models for water and CO$_2$ respectively.
We work at 400 bar and different temperatures and CO$_2$ concentrations. 
We use brute force molecular dynamics when the supersaturation or the supercooling are so high so that nucleation occurs
spontaneously and Seeding otherwise. We used both methods for a particular state 
and found an excellent agreement when using a linear combination of $\overline{q}_{3}$ and $\overline{q}_{12}$ order parameters to identify critical clusters.
 With such order
parameter we get a rate of 10$^{25}\,\text{m}^{-3}\text{s}^{-1}$ for nucleation in a CO$_2$ saturated solution at $255\,\text{K}$ ($35\,\text{K}$ of supercooling).
By comparison with our previous work on methane hydrates, we conclude that nucleation of CO$_2$ hydrates is several orders of magnitude faster due
to a lower interfacial free energy between the crystal and the solution. By combining our 
nucleation studies with a recent calculation of the hydrate-solution interfacial free energy at coexistence [Algaba \emph{et al.}, J.~Colloid Interface Sci.~\textbf{623}, 354–367 (2022)],
we obtain a prediction of the nucleation rate temperature dependence for CO$_{2}$-saturated solutions (the experimentally
relevant concentration). 
On the one hand, we open the window for comparison with experiments for supercooling
larger than $25\,\text{K}$. 
On the other hand, we conclude that homogeneous nucleation is impossible for supercooling lower than
$20\,\text{K}$. Therefore, nucleation must be heterogeneous in typical experiments where hydrate formation is observed at
low supercooling. To assess the hypothesis that nucleation occurs at the solution-CO$_2$ interface we run spontaneous
nucleation simulations in two-phase systems and find, by comparison with
single-phase simulations, that the interface does not affect hydrate nucleation, at least at the deep supercooling at which this study was carried out
($40$ and $45\,\text{K}$). Overall, our work sheds light on molecular and thermodynamic aspects of hydrate nucleation.

\end{abstract}

\maketitle
$^*$Corresponding author: felipe@uhu.es

%

\section{Introduction}


When a liquid is cooled below the solid-liquid coexistence temperature, the crystallization is not an immediate process and the liquid can remain in a metastable supercooled state for some time. Fluctuations still exist and the formation of small embryos of the stable crystal phase can be observed. When these fluctuations lead to the formation of a solid cluster that surpasses a critical size then crystallization cannot be avoided. 
This mechanism is usually known as homogeneous nucleation. In the proximity of the equilibrium freezing temperature, the critical cluster size is quite large and the liquid phase can remain stable for quite a long time. The presence of solid impurities reduces the size of the critical cluster and makes nucleation easier, leading to heterogeneous nucleation that can be observed easily even for temperatures moderately below the freezing temperature.~\cite{debenedetti2020metastable} 

An interesting observable is the nucleation rate $J$ defined as the number of critical clusters per unit of time and volume. The nucleation rate can be determined in experiments, mainly for ice in supercooled water\cite{stan2009microfluidic, taborek1985nucleation, demott1990freezing, stockel2005rates, kramer1999homogeneous, duft2004laboratory, laksmono2015anomalous, manka2012freezing, hagen1981homogeneous, miller1983homogeneous} but only (due to limitations in system size and accessible time) when its value is smaller than $10^{16}/(\text{m}^3\,\text{s})$.
In simulations, the nucleation rate can be determined in brute force (BF) simulations only when its value is of the order of $10^{30}/(\text{m}^3\,\text{s})$ or higher (due to limitations in system size and accessible time). 
Thus, there is a range of nucleation rates between $10^{16}-10^{30}/(\text{m}^3\,\text{s})$ that cannot be accessed either by experiments or by BF simulations. 
However, the use of special rare event technique simulations allows to determine the nucleation rate in this intermediate regime or even for temperatures accessible in experiments.

Several techniques have been proposed to obtain nucleation rates in simulations when BF simulations are not sufficient. Two of them:  Umbrella Sampling\cite{torrie1977nonphysical} (US) and Metadynamics\cite{laio2002escaping} are aimed at determining the free energy barrier for nucleation and the nucleation rate using the formalism proposed by Volmer and Weber~\cite{volmer1926keimbildung} and Becker and D\"oring.~\cite{becker1935kinetische} About 25 years ago Bolhuis and coworkers proposed a methodology, the Transition Path Sampling (TPS),\cite{Bolhuis2002} where an analysis of the trajectories that are reactive (i.e., leading from the metastable phase to the stable phase) is performed, allowing the determination of nucleation rates. About twenty years ago another method, the Forward Flux Sampling (FFS),\cite{Bi2014a,haji2015direct} was proposed to analyze the fraction of successful trajectories leading from one value of the order parameter to the next and the flux to the initial lowest order value of the order parameter considered. 
The nucleation rates obtained by these four methods (Umbrella Sampling, Metadynamics, Transition Path Sampling, and Forward Flux Sampling) are in principle exact (or almost exact) for the considered potential model. 

More recently some of us\cite{Sanz2013a} and independently Knott et \textit{al}.\cite{Knott2012a} introduced a new approximate technique to determine nucleation rates known as Seeding. In this technique, a solid cluster is inserted into a 
metastable fluid and the conditions at which this cluster is critical (i.e., with 50\% probability of evolving to either phase) are determined. This followed the first ideas about using seeds for nucleation studies introduced by Bai and Li.\cite{Bai2005, Bai2006a} Once the size of the critical cluster is determined then the expression of the Classical Nucleation Theory (CNT) is used to estimate the nucleation rate. The main disadvantage of Seeding is that it is an approximate technique as the results depend on the choice of the order parameter. However, its main advantage is its simplicity thus, allowing to study really complex systems for which more rigorous methods are too expensive from a computational point of view. 
It has been shown, that with appropriate order parameters, Seeding correctly predicts the 
nucleation rates of hard spheres, Lennard-Jones systems,\cite{Espinosa2016c} electrolytes\cite{Espinosa2015a} or even the nucleation of ice both from pure water and from aqueous electrolyte solutions.\cite{espinosa2014homogeneous, Soria2018a} Recently we have shown that it can also predict the nucleation rate of hydrate formation for the methane hydrate.\cite{Grabowska2022b}

Hydrates are non-stoichiometric solids formed when a gas (typically methane or carbon dioxide) is in contact with water under moderate pressure (i.e., $30-1000\,\text{bar}$) and the system is cooled. In the most common hydrate structure (sI) the unit cell has cubic 
symmetry and contains 46 molecules of water and 8 molecules of guest (occupying two types of cavities, six large and two somewhat smaller).\cite{Sloan2008a,Ripmeester2022a,Zhang2022a} Methane hydrates can be found naturally on the seafloor near the coasts and it is also formed in the pipes transporting natural gas.~\cite{Ripmeester2016a} It is also expected to be found on some planets.~\cite{Pellenbarg2003a} Although the methane hydrate is the most relevant, the interest in the hydrate containing carbon dioxide (CO$_{2}$) is growing. 
This is so because replacing methane by CO$_{2}$ in the hydrate would be a simple procedure to sequestrate CO$_{2}$ from the atmosphere and to mitigate its greenhouse effect that leads to global warming.~\cite{English2015a,Tanaka2023a}  
    
When the gas is in contact with water the formation of the hydrate starts at a certain temperature denoted as $T_3$.\cite{Sloan2008a} This temperature 
is indeed a triple point, where three phases: the solid hydrate, the aqueous solution, and the gas, coexist at equilibrium. The value of $T_3$ depends on the pressure, and nucleation rates increase dramatically as one moves from $T_3$ to lower temperatures at constant pressure.

Several experimental studies deal with hydrate nucleation.~\cite{Ruoff1994, Myerson1999, Uchida2000, Cournil2004, Svartaas2010, Solms2018, Maeda2018, Maeda2019} Many computational studies on hydrate nucleation have also been reported.~\cite{Clancy1994, Smith1996, Alavi2005a,Alavi2006a,Wu2009, English2009a,Walsh2011a, Sarupria2012b, Knott2012a, Liang2013a, Barnes2014a, Bi2014a, Yuhara2015a, Zhang2016b, Lauricella2017a, Arjun2019a, Arjun2020a, Arjun2021a, Arjun2023,Liang2013a,Zhang2018a,Zhang2020a,Wang2003a,Jimenez-Angeles2014a,Jimenez-Angeles2018a,Zhang2022a,Tanaka2023b,Tanaka2024a} Comparison between experimental and simulation studies is difficult due to the presence of heterogeneous nucleation in experiments, and to the fact that in many simulation studies one must use large supersaturations (i.e., solubilities of the gas artificially higher than the experimental ones) to increase the driving force to facilitate the kinetics of the nucleation process. In their pioneering molecular dynamics study, Walsh et \textit{al.}\cite{Wu2009} used a high concentration of methane in the aqueous solution in order to observe nucleation events in a reasonable simulation time. In fact, using a supersaturated solution of the guest molecule is a common strategy in the studies of nucleation of hydrates. There are two ways to prepare such a system. The first one is to use a homogeneous solution of guest molecule in water,\cite{Sarupria2012b, Liang2013a} in which the concentration of solute is higher than the equilibrium solubility under the same conditions. However, this is only possible at low temperatures, where the nucleation of hydrate is faster than the nucleation of gas bubbles.~\cite{Grabowska2022a} The second option is to use a system in which there is a curved interface between the solution and a gas phase\cite{Wu2009, Arjun2019a, Arjun2020a, Arjun2021a} (i.e., bubbles of gas in the solution). The presence of a curved interface results in an increase in the solubility of the guest molecule in water. These two methods allow to obtain spontaneous nucleation events in BF simulations.~\cite{Sarupria2012b, Liang2013a, Wu2009} Additionally, Arjun \emph{et~al.}~\cite{Arjun2019a, Arjun2020a, Arjun2021a} were able to estimate the nucleation rate of hydrates at temperatures well below the $T_3$ by combining transition path sampling with the use of gas bubbles, that increases the solubility of the gas. In experiments, however, the concentration of guest molecules in the solution is dictated by the equilibrium solubility of the solute in water \textit{via} a planar interface. For that reason, in this work we study the nucleation of hydrate under experimental conditions, i.e., we use the concentration of guest molecule (CO$_{2}$ in this work) equal to its equilibrium solubility. To the best of our knowledge, there are only two simulation studies where the nucleation rate was computed under "realistic experimental conditions" (without supersaturation) for the formation of the methane hydrate. In the first one, Arjun and Bolhuis \cite{Arjun2023} in a tour de force used TPS to determine the nucleation rate. In the second one, we used the Seeding technique.~\cite{Grabowska2022b}  Good agreement was found between the estimates of the nucleation rate from these two studies.

In this work, we shall use the Seeding technique\cite{Espinosa2016c} to determine by computer simulations the homogeneous nucleation rate of the CO$_{2}$ hydrate at the pressure of  400 bar and when the supercooling
$\Delta T=T_3-T$, (i.e., the difference between the dissociation temperature $T_{3}$ and the current temperature $T$) is equal to 35 K.  We shall determine the nucleation rate under experimental conditions (i.e., without supersaturation). This study is a follow-up of a previous study where we used the same technique to study the nucleation rate of methane hydrate at the same pressure and degree of supercooling \cite{Grabowska2022b}. The comparison will be useful as it illustrates the differences in the nucleation rate of hydrates of methane and CO$_{2}$ at the same thermodynamic conditions (i.e., equal pressure and degree of supercooling). At first, one would expect that the differences between both gases should not be too large as the guest molecules are of similar size. However, the solubility of CO$_{2}$ in water is an order of magnitude larger than that of methane (due to its large quadrupole moment leading to more favorable water-gas interactions).  It will be shown that the nucleation rate of CO$_{2}$ hydrate is much higher for a certain fixed pressure and a certain fixed supercooling compared to methane hydrate. The comparison is especially useful as we are using the same water model that was employed in our previous study of methane, namely TIP4P/Ice. Although the higher nucleation rate for the CO$_{2}$ hydrate may be due to its higher solubility,~\cite{Zhang2018b}  we think the main reason for this is the lower value of the interfacial free energy between the hydrate and the aqueous solution.


Finally, we shall analyze the impact of the gas-water interface on the nucleation rate.
Hydrates are always obtained in experiments by considering a two-phase system (gas in contact with water). There is the possibility that the presence of the interface facilitates the nucleation of the solid phase. Thus, heterogeneous nucleation 
(due to the presence of the gas-water interface rather than to the presence of solid impurities in the liquid phase) may be responsible for the nucleation found in experiments. To determine this point we performed simulations both in the presence and in the absence of the interface with the same concentration of CO$_{2}$ in aqueous solution. We conclude that nucleation rates obtained in both cases were the same, suggesting that the gas-water interface does not enhance the nucleation rate, at least for the thermodynamic conditions considered in this work. 
       
The organization of this paper is as follows. In Sec. II, we describe the methodology used in this work. The results obtained, as well as their discussion, are described in Sec. III. Finally, conclusions are presented in Sec. IV.

\section{Methodology}

\subsection{Seeding: A brief description}

From the description of the CNT,\cite{volmer1926keimbildung, farkas1927keimbildungsgeschwindigkeit, becker1935kinetische, zeldovich1943theory} the formation of a solid cluster of size $N$ at given temperature $T$ and pressure $P$,  into the liquid phase requires a free energy of formation $\Delta G$ given by:

\begin{equation}
    \Delta G=-N\,|\Delta \mu_{\text{N}}| + \gamma \, \mathcal{A}
\end{equation}

\noindent
where $\Delta\mu_{\text{N}}$ is the driving force for nucleation. In the case of a pure substance, it is just the difference in chemical potentials of the solid and fluid phases at the considered thermodynamic conditions. In the case of hydrate formation, it is simply the difference between the chemical potential of the solid phase and that of the hydrate molecules in the liquid phase (we shall come to this point later). $\gamma$ is the solid-liquid interfacial free energy, and $\mathcal{A}$ is the interfacial area. Since the first term is negative and grows with $N$ and the second is positive and grows with the area (i.e., $N^{2/3}$) a maximum is reached for a certain value of $N$ (i.e., the size of the critical cluster $N_c$) leading to a free energy barrier of $\Delta G_c$: 

\begin{equation}
\Delta G_{c} = \frac{1}{2}N_{c} \, |\Delta \mu_{\text{N}}| \,
    \label{driving_force_cnt}
\end{equation}

The size of the critical cluster can be obtained as:
\begin{equation}
    N_{c}=\frac{32\,\pi\,\gamma^{3}}{3\,\rho_{s}^{2}\,|\Delta \mu_{\text{N}}|^{3}}
    \label{critical_nucleus_cnt}
\end{equation}
where $\rho_s$ is the number density of the bulk solid phase at the considered $P$ and $T$ of the system (in CNT one neglects changes in the density of the solid in the critical cluster due to the Laplace pressure which is equivalent to assume that the solid phase is incompressible). The free energy barrier can also be rewritten as 

\begin{equation}
   \Delta G_{c}= \frac{16\,\pi\,\gamma^{3}}{3\,\rho_{s}^{2}\,|\Delta \mu_{\text{N}}|^{2}}
   \label{eq:barrier}
\end{equation}

According to CNT, if a steady state is considered, i.e., the distribution of clusters of different sizes does not depend on time,
the nucleation rate per unit volume $J$ at a given temperature $T$ is the product of the probability of a critical nucleus formation, which depends on the free energy of formation $\Delta G_{c}$ as $\mathcal{P}(N_{c})\approx e^{-\Delta G_{c}/k_{B}T}$ and a kinetic factor $J_{0}$:

\begin{equation}
    J = J_{0} \, e^{- \Delta G_{c}/k_{B}T} = \rho_{f} \, Z \, f^{+} \, e^{-\Delta G_{c}/k_{B}T}
     \label{nucleation_rate_cnt_1}
\end{equation}

\noindent where $k_{B}$ is the Boltzmann constant and the $J_{0}=\rho_{f} \, Z \, f^{+}$ term contains the kinetic growth information through the fluid number density $\rho_{f}$. $Z$ is the the Zeldovich factor which is given by:
\begin{equation}
Z = \sqrt{\frac{|\Delta G_{c}''|}{2\pi k_{B}T}} =\sqrt{\frac{|\Delta \mu_{\text{N}} |}{ 6 \,\pi \, k_{B}T N_{c} }}
    \label{zeldovich}
\end{equation}

\noindent 
Here $\Delta G_{c}''$ is the curvature of the free energy formation at the critical size.

The attachment rate $f^{+}$ which can be calculated via an effective diffusion constant that accounts for the number of particles aggregated and separated in time from the critical cluster as follows:

\begin{equation}
f^{+} = \frac{ \left\langle \Delta N_{c}^{2} (t) \right \rangle }{2 \, t} = \frac{ \left\langle \left[ N_{c} (t) - N_{c}(t_{0})\right]^{2} \right \rangle }{2 \, t}
    \label{attachment_rate}
\end{equation}
\\

We have shown in a previous work\cite{Grabowska2022b} that the corresponding expression of CNT for the hydrate nucleation can be written as:

\begin{equation}
J = \rho_{L}^{\text{CO}_{2}} \,   Z \, f^{+}_{\text{CO}_{2}} \, \text{exp} \left(\frac{-N_{c}^{\text{CO}_{2}} \, |\Delta \mu_{\text{N}} | }{2 \, k_{B} \, T } \right) 
    \label{nucleation_rate_cnt_2}
\end{equation}

\noindent
where $\rho^{\text{CO}_{2}}_{L}$ is the number density of CO$_{2}$ in the liquid phase, $N_c^{\text{CO}_{2}}$ is the number of molecules of CO$_{2}$ in the critical cluster (notice that the critical cluster contains both molecules of water and molecules of CO$_{2}$) and $f^{+}_{\text{CO}_2}$ is the attachment rate computed from Eq.~\eqref{attachment_rate} by analyzing the diffusive behavior or the number of CO$_{2}$ molecules in the solid cluster when starting from configurations at the critical size. The value of $\gamma$ can be obtained from Eq.\eqref{critical_nucleus_cnt} by using 
$\rho_{S}^{\text{CO}_{2}}$ which is just the number density of molecules of CO$_{2}$ in the hydrate. 

In the Seeding technique, a solid cluster is inserted into the metastable fluid at the thermodynamic conditions at which it is
critical (i.e., $50\%$ of probability of either melting or growing is determined). Once the size of the critical cluster $N_c$ (where $N_{c}$ is the number of CO$_{2}$ molecules in the solid critical cluster)  is known one determines the free energy barrier using Eq.~\eqref{driving_force_cnt}. The value of  
$\rho_{L}^{\text{CO}_{2}}$ is determined from the solubility of CO$_{2}$ at the considered value of $P$ and $T$ (or with a higher value in the case of supersaturated solutions as it will be shown later on). 
The only remaining ingredient is $\Delta \mu_{\text{N}}$ which will be described in detail in the next subsection.

\subsection{Driving Force for nucleation}

$\Delta \mu_{\text{N}}$ can be viewed, as first suggested by Kashchiev and Firoozabadi~\cite{Kashchiev2002a} (see also our previous works~\cite{Grabowska2022a,Grabowska2022b,Algaba2023a}),  as a chemical reaction that takes place at constant $P$ and $T$.
In fact  $\Delta \mu_{\text{N}}$ is just the chemical potential change of the following physical process: 

\begin{equation}
\text{CO}_{2} (\text{aq},x_{\text{CO}_{2}}) +
5.75\,\text{H}_{2}\text{O} (\text{aq},x_{\text{CO}_{2}}) 
\rightarrow [\text{CO}_{2}(\text{H}_{2}\text{O})_{5.75}]_{\text{H}}
\label{reaction}
\end{equation}

\noindent
In Eq.~\eqref{reaction}, one molecule of CO$_{2}$ in the aqueous solution reacts with $5.75$ molecules of water (also in the aqueous solution phase) to form a $[\text{CO}_{2}(\text{H}_{2}\text{O})_{5.75}]_{\text{H}}$ ``hydrate molecule'' in the solid phase.  Since we are assuming, as in our previous works,~\cite{Grabowska2022a,Grabowska2022b,Algaba2023a} that all cages of the hydrates are filled. A unit cell of CO$_{2}$ hydrate is formed by $46$ water molecules and $8$ CO$_{2}$ molecules, i.e., one molecule of  CO$_{2}$ reacts with  $46/8=5.75$ water molecules. This is consistent with the stoichiometric reaction given by Eq.~\eqref{reaction}.

Since all cages of the hydrate are occupied, the chemical potential of this compound (hydrate) can be obtained as the sum of the chemical potential of CO$_{2}$ in the solid plus $5.75$ times the chemical potential of water in the solid (see Eq.~(8) of our previous paper~\cite{Algaba2023a}).  Note that the chemical potentials depend on $T$ and $P$ (and on composition). However, all the results of this work were obtained for a pressure of $400\,\text{bar}$. For this reason, we shall omit the pressure dependence and will write the chemical potential of the hydrate simply as $\mu_{\text{H}}^{\text{H}}(T)$ (there is no dependence on composition for the hydrate as its stoichiometry is fixed). Following the work of Kashchiev and Firoozabadi~\cite{Kashchiev2002a} and our previous works,~\cite{Grabowska2022a,Grabowska2022b,Algaba2023a} the driving force for nucleation of the hydrate formed from the aqueous solution with a concentration $x_{\text{CO}_{2}}$ at $T$ can be expressed as:
\begin{align}
\Delta\mu_{\text{N}}(T,x_{\text{CO}_{2}})&=\mu^{\text{H}}_{\text{H}}(T) \nonumber\\
& -\mu^{\text{aq}}_{\text{CO}_{2}}(T,x_{\text{CO}_{2}})-5.75\,\mu^{\text{aq}}_{\text{H}_{2}\text{O}}(T,x_{\text{CO}_{2}})
\label{driving_force}
\end{align}

\noindent where $\mu^{\text{aq}}_{\text{CO}_{2}}(T,x_{\text{CO}_{2}})$ is the chemical potential of CO$_{2}$ in the aqueous solution, and  $\mu^{\text{aq}}_{\text{H}_{2}\text{O}}(T,x_{\text{CO}_{2}})$ is the chemical potential of water in the
aqueous solution.

The nucleation rate of the CO$_{2}$ hydrate has been determined by using BF simulations for most of the cases. In BF runs $J$ is determined directly and it is not necessary to know the value of $\Delta \mu_{\text{N}}$. However, for two thermodynamic states, it was necessary to use the Seeding method, and therefore it was necessary to obtain the value of
$\Delta \mu_{\text{N}}$ to determine the nucleation rate. In this context, it is useful to introduce the supersaturation at a given pressure $P$ and temperature $T$ defined as:

\begin{equation}
S=\dfrac{x_{\text{CO}_{2}}}{x^{\text{eq}}_{\text{CO}_{2}}}
\label{supersaturation}
\end{equation}

\noindent
where $x_{\text{CO}_{2}}$ is the CO$_{2}$ molar fraction of a solution and $x^{\text{eq}}_{\text{CO}_{2}}$ is the CO$_{2}$ molar fraction at experimental conditions, i.e., the CO$_{2}$ concentration in water when in equilibrium with pure CO$_{2}$ via a planar interface at the same $P$ and $T$. In particular, we used the Seeding method for $T=255\,\text{K}$ when $S=1$ and when $S=1.207$. Note that $S=1$ is the setup used in experimental work.  We shall also determine $J$ from Seeding at $255\,\text{K}$ for the case $S=1.207$. This case is interesting as for this particular state it is possible 
to determine $J$ both from BF runs and from the Seeding method and this state serves as a cross-check of the Seeding method (in particular of the choice of the order parameter).  The states for which we determined $J$ in this work are shown in Fig.~\ref{fig:pd} as diamonds and triangles.

Since Seeding is used here only for two thermodynamic states at $255\,\text{K}$ and $400\,\text{bar}$, namely  $S=1$  and $S=1.207$ only the values of $\Delta \mu_{\text{N}}$ for these two states are needed. 
In our previous work,\cite{Algaba2023a} the driving force for nucleation of the hydrate of CO$_2$ has been obtained using four independent methods along the solubility curve obtained when a CO$_2$-rich phase is in contact with an aqueous phase at $400\,\text{bar}$ and several temperatures below the dissociation temperature. Particularly, we have proposed a novel methodology to evaluate the driving force for nucleation based on the calculation of partial enthalpies of CO$_{2}$ and water in the aqueous phase at different values of CO$_{2}$ composition and temperatures (we recommend to read Section E.4 of our previous work\cite{Algaba2023a} for further details). This is a rigorous methodology obtained only from thermodynamic arguments for calculating the driving force for nucleation of the CO$_{2}$ hydrate at any $P$, $T$ and $x_{\text{CO}_{2}}$. According to this, it is possible to directly determine the value of the driving force for nucleation at $255\,\text{K}$ and $400\,\text{bar}$, at the equilibrium solubility composition (i.e., $S=1$ or $x_{\text{CO}_{2}}=x_{\text{CO}_{2}}^{eq}=0.0803$) when a CO$_2$-rich phase is in contact with an aqueous phase via a planar interface, being it of -2.26 (in $k_{B}T$ units) at $S=1$ and -2.73 (in $k_{B}T$ units) for the case $S=1.207$. See the work of Algaba \emph{et al.}~\cite{Algaba2023a} for further details, and more specifically the route 4 for calculating the driving force (Eq.~(26) in that paper) and Fig.~14 (also there) from which these values are extrapolated.

  
\subsection{Simulation details}

All Molecular Dynamics (MD) simulations are performed using the GROMACS package.~\cite{VanDerSpoel2005a,hess2008gromacs} We use the Verlet leapfrog algorithm\cite{Cuendet2007a} with a time step of $2\,\text{fs}$. The temperature is kept constant using the Nos\'e-Hoover thermostat with a relaxation time of $2\,\text{ps}$.~\cite{Nose1984a,Hoover1985a} The pressure is also kept constant but using the Parrinello-Rahman barostat~\cite{Parrinello1981a} with the same relaxation time. We use two different versions of the $NPT$ or isothermal-isobaric ensemble. For BF simulations and Seeding simulations at supersaturated conditions, we use the isotropic $NPT$ ensemble, i.e., the three sides of the simulation box are changed proportionally to keep the pressure constant. For Seeding simulations at experimental conditions, i.e., at coexistence conditions at which the aqueous solution of CO$_{2}$ and the CO$_{2}$-rich liquid phase coexist, we use the anisotropic $NP_{z}\mathcal{A}T$ ensemble since a planar liquid-liquid interface exists and only fluctuations of the volume are performed varying the length of the simulation box along the $z$-axis direction, perpendicular to the planar interface. We use a cutoff distance of $1\,\text{nm}$ for dispersive and Coulombic interactions. For electrostatic interactions, we use the Particle Mesh Ewald (PME) method.\cite{Essmann1995a} We do not use long-range corrections for dispersive interactions. Water and CO$_{2}$ molecules are described using the TIP4P/Ice\cite{Abascal2005b} and TraPPE\cite{Potoff2001a} models, respectively. TIP4P/Ice predicts correctly the melting point of ice Ih and that guarantees good predictions for the phase equilibria of hydrates.\cite{Conde2013a} The water-CO$_{2}$ unlike dispersive interactions are taken into account via the modified Berthelot rule proposed by M\'{\i}guez \emph{et al.}~\cite{Miguez2015a} and also used in our previous work.~\cite{Algaba2023a} This strategy allows us to predict accurately the three-phase CO$_{2}$ hydrate-water-CO$_{2}$ coexistence or dissociation line of the CO$_{2}$ hydrate. Particularly, with this choice the dissociation temperature or $T_{3}$, at $400\,\text{bar}$, is in excellent agreement with experimental data taken from the literature (see Fig.~10 and Table~II of the work of M\'{\i}guez \emph{et al.}~\cite{Miguez2015a} for further details). It is also important to mention that the same molecular parameters can accurately predict the CO$_{2}$ hydrate–water interfacial free energy.~\cite{Algaba2022b,Zeron2022a,Romero-Guzman2023a}

The dissociation temperature or $T_{3}$ of the CO$_{2}$ hydrate at $400\,\text{bar}$ is $290\,\text{K}$~\cite{Algaba2023a} (close to the experimental value at this pressure which is $286\,\text{K}$).  In this work, all simulations are carried out at $245$, $250$, and $255\,\text{K}$ (supercoolings of $45$, $40$, and $35\,\text{K}$, respectively). Following our previous work,~\cite{Grabowska2022b} we perform three different kinds of simulations to determine the nucleation rate of the CO$_{2}$ hydrate at $255\,\text{K}$: (1) BF simulations at supersaturation conditions; (2) Seeding simulations at supersaturation conditions; and (3) Seeding simulations at experimental saturated conditions. In the first set of simulations, we estimate the nucleation rate of the hydrate at two different saturated conditions using its definition and the mean first-passage time approach (MFPT). In the second set, we also determine the nucleation rate at one of the supersaturation conditions following the Seeding approach. This allows us to check if the local bond order parameters used to characterize the size of the critical cluster of the hydrate are appropriate. Finally, in the third set of simulations and once we have got the best selection of the order parameters, we estimate the nucleation rate of the CO$_{2}$ hydrate at experimental conditions, i.e., at the equilibrium (saturated) conditions of CO$_{2}$ in water in contact with a CO$_{2}$-rich liquid phase via a planar interface using the Seeding Technique.

\begin{figure}
\centering
\hspace*{-0.2cm}
\includegraphics[width=\columnwidth]{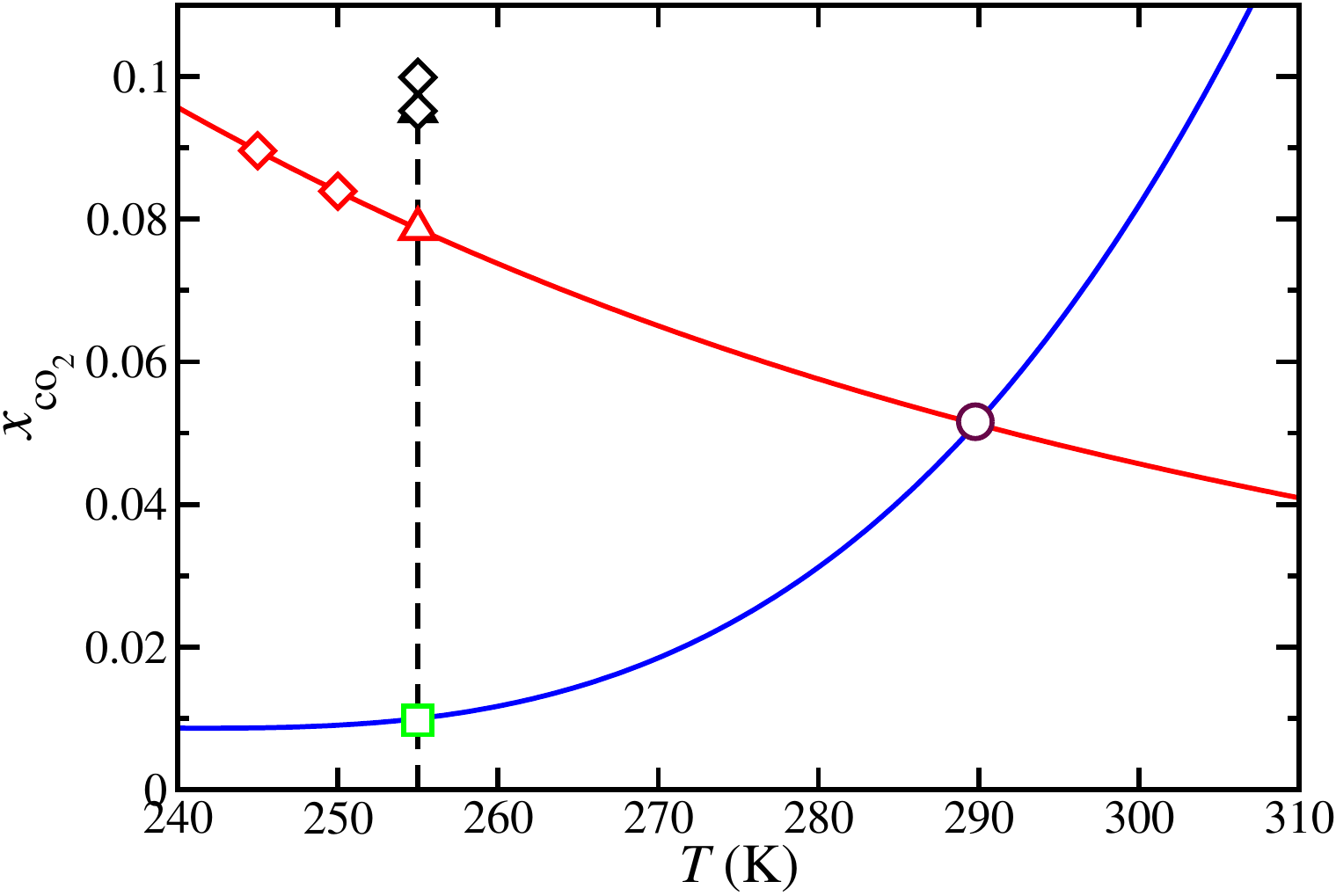}

\caption{CO$_2$ molar fraction, $x_{\text{CO}_2}$, in an aqueous solution coexisting with the hydrate (blue curve) and with a CO$_2$ fluid reservoir (red curve), as functions of temperature, at $400\,\text{bar}$. 
Diamonds and triangles represent the six states at which the nucleation rate of CO$_{2}$ hydrate, $J$, is obtained in this work using BF simulations (diamonds) and the Seeding technique (triangles). Red color symbols are used to denote saturated conditions ($S=1$) and black color symbols to denote supersaturated conditions ($S>1$). Note that at $255\,\text{K}$ and $S=1.207$ ($x_{\text{CO}_{2}}=0.0969$) we have estimated $J$ from BF (black diamond) and Seeding simulations (black triangle). The crossing point between both curves (maroon circle) corresponds to the temperature $T_3$  at which hydrate, solution, and CO$_2$ coexist.~\cite{Tanaka2018a,Grabowska2022a} The green square represents the hydrate-solution coexistence point at $255\,\text{K}$.}
\label{fig:pd}
\end{figure}

\begin{figure}
\centering
\hspace*{-0.2cm}
\includegraphics[width=0.5\columnwidth]{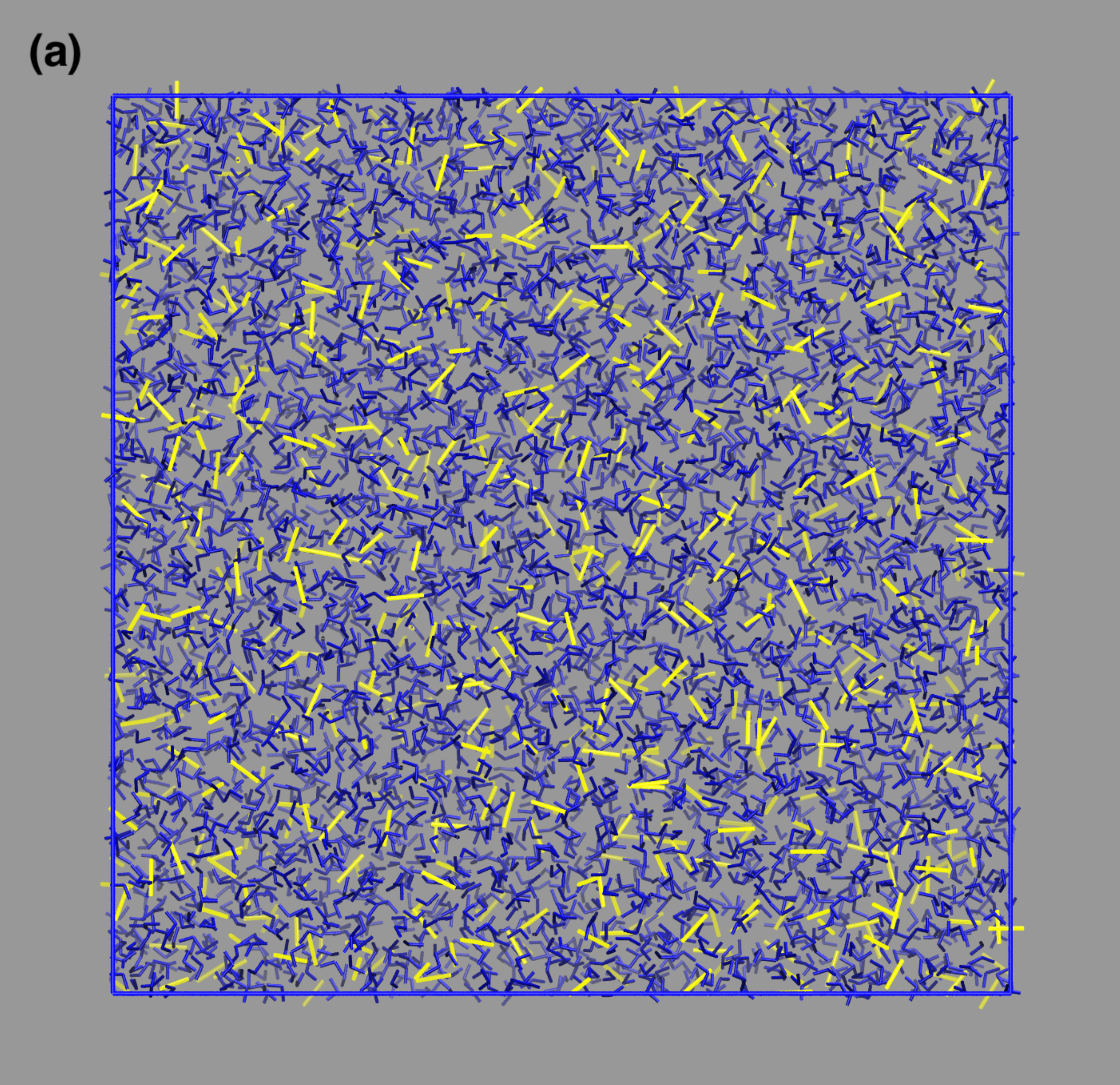}
\includegraphics[width=0.5\columnwidth]{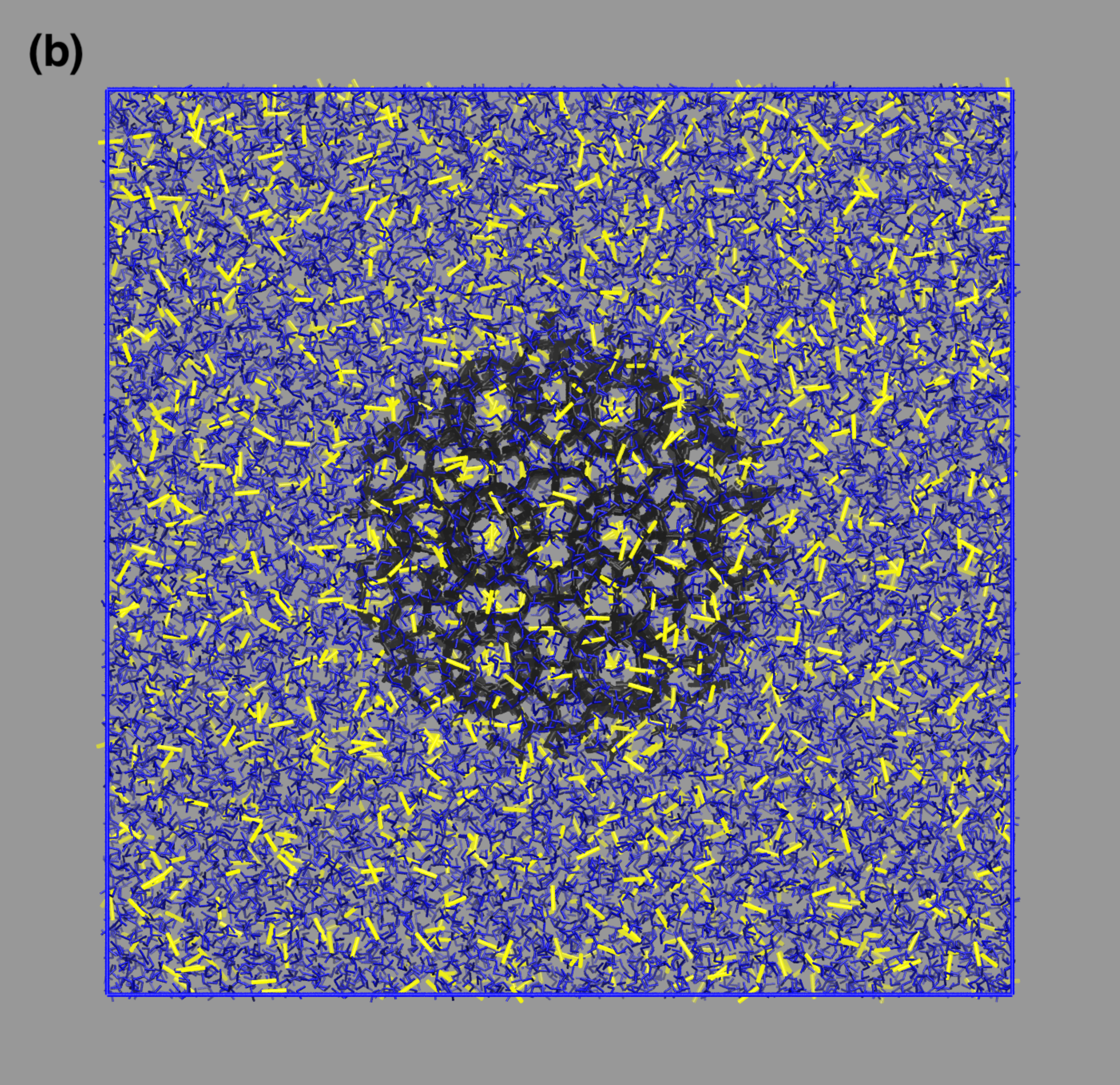}
\hspace*{-0.2cm}
\includegraphics[width=1.0\columnwidth]{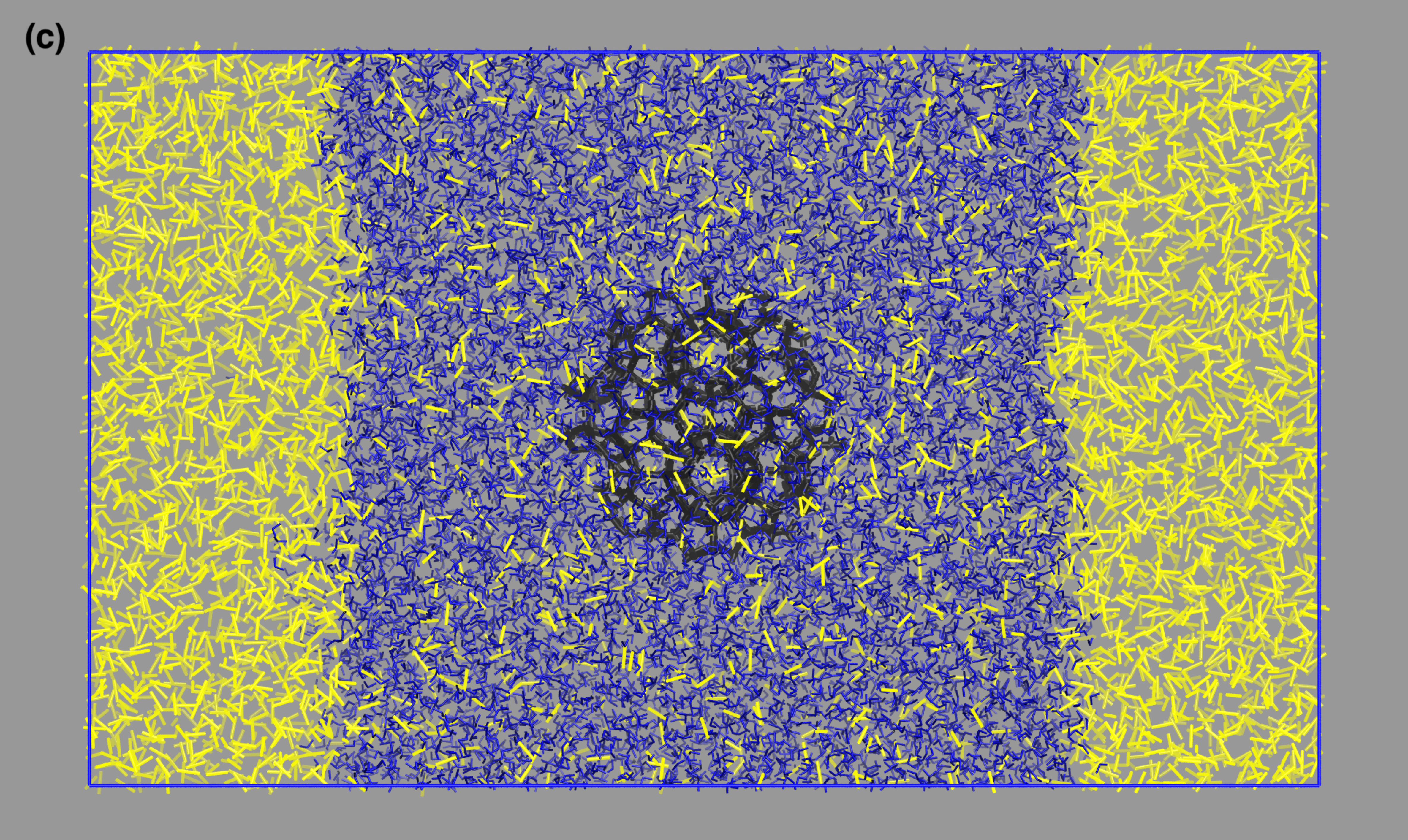}
\caption{Starting configurations for runs performed in this work to estimate nucleation rates at $400\,\text{bar}$ and different temperatures and concentrations. Water and CO$_{2}$ molecules are represented as blue and yellow sticks, respectively. Black molecules depict the spherical seed of CO$_{2}$ hydrate introduced into the system to induce crystallization. (a) One-phase system of a supersaturated aqueous solution of CO$_{2}$ ($5.57\times5.57\times5.57\,\text{nm}^{3}$). (b) Two-phase system with a solid cluster of CO$_{2}$ hydrate (seed) inserted in the aqueous solution of CO$_{2}$ at saturation $S>1$ ($5.58\times5.58\times5.58\,\text{nm}^{3}$). (c) Three-phase system with a CO$_{2}$ liquid phase in contact with an aqueous solution of CO$_{2}$ via a planar interface and a spherical cluster of CO$_{2}$ hydrate inserted in the aqueous solution of CO$_{2}$ ($7.40\times7.40\times12.41\,\text{nm}^{3}$). In this case, the concentration of CO$_{2}$ in water is that of equilibrium ($S=1$). The size of the simulation boxes is given in terms of average values since it fluctuates in $NPT$ simulations. In case (c), simulations are performed in the $NPz\mathcal{A}T$ ensemble so that $L_{x}$ and $L_{y}$ are fixed and $L_{z}$ fluctuates around the average value.}
\label{figure2}
\end{figure}

We perform BF simulations in the isotropic $NPT$ ensemble at $255\,\text{K}$ placing $4942$ water molecules and $530$ and $560$ CO$_{2}$ molecules, respectively (i.e.,$x_{\text{CO}_{2}}=0.0969$ and $x_{\text{CO}_{2}}=0.1018$ ) in a cubic 
simulation box as shown in Fig.~\ref{figure2}a. 
As  the concentration of CO$_{2}$ in water at coexistence conditions is $x^{\text{eq}}_{\text{CO}_{2}}=0.0803$,\cite{Algaba2023a} the systems considered correspond to $S=1.207$ and $S=1.268$ respectively. In all cases, the system is equilibrated during $5-10\,\text{ns}$ and run during up to $3\,\mu\text{s}$. This simulation time allows us to observe the formation of solid clusters of the CO$_{2}$ hydrate. We show in Fig.~\ref{fig:pd} some of the states for which we determined the nucleation rate at $255\,\text{K}$. The states simulated using BF simulations at this temperature are represented as black triangles. In our previous works, we have determined the dissociation temperature $T_{3}$ of the CO$_{2}$ and CH$_{4}$ hydrates using the so-called solubility method and calculating the crossing point (maroon circle) between the solubility curves of CO$_{2}$ in the aqueous solution when it is in contact with the CO$_{2}$ liquid phase and the hydrate, as it is shown in Fig.~\ref{fig:pd}. This methodology has also been used in previous work by Tanaka and coworkers (see Fig.~9 of their work).~\cite{Tanaka2018a}

We also perform Seeding simulations at one of the two supersaturated concentrations, $S=1.207$. According to the Seeding technique, a spherical cluster of CO$_{2}$ hydrate is inserted into a supersaturated aqueous phase of CO$_{2}$ in water, as it is shown in Fig.~\ref{figure2}b. To do this, we first consider two bulk phases, one of CO$_{2}$ hydrate and another of an aqueous solution of CO$_{2}$ with the appropriate supersaturation ($S=1.207$), at $255\,\text{K}$ and $400\,\text{bar}$. The aqueous solution of CO$_{2}$ is identical to that used in the BF simulations ($4942$ water molecules and $530$ CO$_{2}$ molecules). The hydrate simulation box is formed from $2944$ molecules of water and $512$ CO$_{2}$ molecules. This corresponds to a $4 \times 4 \times 4$ unit cell of sI hydrate structure with full occupancy. The space group of the unit cell is ${\displaystyle Pm{\overline {3}}n}$. The proton disorder was obtained using the algorithm of Buch et al.\cite{buch1998simulations} Both simulation boxes are equilibrated in the $NPT$ ensemble separately. The hydrate system is equilibrated during $50\,\text{ns}$. After this time, a spherical cluster of radius ranging from $1$ to $1.4\,\text{nm}$ is cut and immersed into the saturated aqueous solution of CO$_{2}$ in water. This is practically done by removing water and CO$_{2}$ molecules and creating a spherical empty space with the same radius of seed of the spherical hydrate cluster. Overlaps in the interface are avoided by slightly moving or rotating nearby molecules. We recommend the reader our previous work for further details.~\cite{Grabowska2022a}

Additionally, we run Seeding simulations at coexistence conditions, i.e., the hydrate cluster is inserted into a solution in equilibrium with a CO$_{2}$-rich liquid phase via a planar interface at $255\,\text{K}$ and $400\,\text{bar}$. This corresponds to a molar fraction of CO$_{2}$ in water $x_{\text{CO}_{2}}=x^{\text{eq}}_{\text{CO}_{2}}=0.0803$. This state corresponds to the red triangle represented in Fig.~\ref{fig:pd} at $255\,\text{K}$.

To keep this concentration constant, the hydrate seed is inserted into the aqueous solution in contact with a CO$_{2}$-rich liquid phase, as shown in Fig.~\ref{figure2}c. In this case, since there is a planar interface in the simulation box, we perform the simulations in the $NP_{z}\mathcal{A}T$ anisotropic ensemble. The aqueous solution - CO$_{2}$ system is formed from $12000$ water molecules and $4952$ CO$_{2}$ molecules. These are the total number of molecules of the whole system, the aqueous solution of CO$_{2}$ and the CO$_{2}$ liquid reservoir. Once the system is properly equilibrated, the spherical hydrate is inserted in the center of the aqueous phase in the same way as in the Seeding simulations at supersaturated conditions.

Finally, we also perform additional BF simulations at $245$ and $250\,\text{K}$ (at $400\,\text{bar}$ in both cases). In both cases, however, simulations are performed at $S=1.0$. i.e., at the corresponding CO$_{2}$ saturation concentration. 
These two states correspond to the red diamonds represented in Fig.~\ref{fig:pd} at $245$ and $250\,\text{K}$. We use two types of simulation setups for this study: a homogeneous CO$_{2}$ saturated bulk solution and a saturated solution in contact via a planar interface with a fluid CO$_{2}$ reservoir. In the first case, we use isotropic $NPT$ runs. In the second one, we use $NP_{z}\mathcal{A}T$ runs. The reason to determine $J$ in these two different setups is that we want to investigate if the presence of an interface between CO$_{2}$ and water enhances/hinders or has no effect on the nucleation rate. At $245\,\text{K}$, we use $2400$ and $240$ water and CO$_{2}$ molecules in the homogeneous system (cubic simulation box with a volume of $82.5\,\text{nm}^{3}$) and $2400$ and $1148$ water and CO$_{2}$ molecules in the inhomogeneous system (volume of the simulation box equal to $4141.9\,\text{nm}^3$). This corresponds in both cases to a molar fraction of CO$_{2}$ in water $x_{\text{CO}_{2}}=x^{\text{eq}}_{\text{CO}_{2}}=0.09$. At $250\,\text{K}$, we use $6524$ and $606$ water and CO$_{2}$ molecules in the homogeneous system and $7200$ and $3444$ water and CO$_{2}$ molecules in the inhomogeneous system. As in the previous case, in the homogenous system we use a cubic simulation box with a volume of $222.5\,\text{nm}^3$. In the inhomogeneous system, the volume of the simulation box is $420.1\,\text{nm}^3$. In this case, the molar fraction of CO$_{2}$ in water $x_{\text{CO}_{2}}=x^{\text{eq}}_{\text{CO}_{2}}=0.085$.

It is important to recall here that the size of our system, as well as the number of molecules forming the systems in which BF and Seeding simulations are performed at $255\,\text{K}$ (at different supersaturations), have been appropriately selected. Note that we have used the same size for the simulation box and number of water molecules as in our previous work for CH$_{4}$ hydrates.~\cite{Grabowska2022a} The number of CO$_{2}$ molecules is different since the molar fraction in the aqueous solution is different. In BF simulations, when the hydrate cluster size is greater than a threshold ($n_{h}=125$ in this work as is shown in Section III.B), the number of CO$_{2}$ molecules in the cluster is $125/5.75\approx 22$, assuming full occupancy of the hydrate, i.e., $46/8=5.75$ water molecules per each CO$_{2}$ molecule. This means that the molar fraction in the aqueous solution surrounding the hydrate cluster is $0.0954$ and $0.1007$ for $S=1.207$ and $1.268$, respectively. Comparing these values with those at the beginning of the simulations, $0.0969$ and $0.1018$, the variation in $x_{\text{CO}_{2}}$ when a hydrate cluster grows irreversibly is below $1.6\%$. Consequently, we think the concentration of CO$_{2}$ in aqueous solution does not substantially decrease as the hydrate size grows.

In addition, Weijs \emph{et al.}~\cite{Weijs2012a} have reported the existence of a diffusive shielding effect in simulations involving nanobubble clusters that help to stabilize them. We believe there is no diffusive shielding effect in our simulations. The simulation boxes and system sizes used in this work are similar to those employed in our previous work on CH$_{4}$ hydrates,~\cite{Grabowska2022a} where we did not detect such an effect. For instance, in BF simulations with $S=1.207$ performed in this work, the radius of the largest cluster formed from more than $125$ water molecules (threshold value mentioned in the previous paragraph) is lower than $r=1.04\,\text{nm}$. According to this, the minimum distance between any two molecules from the cluster and its periodic image is above $3.5\,\text{nm}$, which corresponds to $3.5\times r_{c}$ with $r_{c}$ the cutoff distance. This confirms that there are no interactions between a cluster and its periodic images. Furthermore, it is worth noting that, within statistical error, one single hydrate cluster is detected in our simulations using the $\overline{q}_{3}-\overline{q}_{12}$ combination of order parameters.

\section{Results}

\subsection{Order parameter}

In general, the size of the largest solid cluster is an adequate order parameter in nucleation studies. To identify the size of the largest solid cluster it is necessary to identify solid and fluid molecules first. A good order parameter should label most of the molecules as fluid when they are in the bulk fluid phase or as solid when they are in the bulk solid phase. The mislabeling (i.e., molecules labeled as solid in the bulk fluid and as liquid in the bulk solid) should be as small as possible
and equal in both phases \cite{Espinosa2016c}. To identify water solid particles we use the averaged order parameters proposed by Lechner and Dellago.~\cite{Lechner2008a} In previous works we have shown that $\overline{q}_{12}$ does a very good job in identifying water molecules in the solid CH$_{4}$ hydrate~\cite{Grabowska2022b} but also in other hydrates.~\cite{Zeron2024b} Here we shall use a combination of  $\overline{q}_{3}$ and $\overline{q}_{12}$ since it provides even better results. Oxygen atoms (and not hydrogen ones) were used when computing the order parameter.  To obtain 
either $\overline{q}_{3}$ or  $\overline{q}_{12}$ of each water molecule we considered all the water molecules at a distance of $5.5\,\text{\AA}$ or less from the molecule of interest (this distance corresponds to the second minimum of the radial distribution function). 

\begin{figure}
\hspace*{-0.2cm}
\includegraphics[width=0.9\columnwidth]{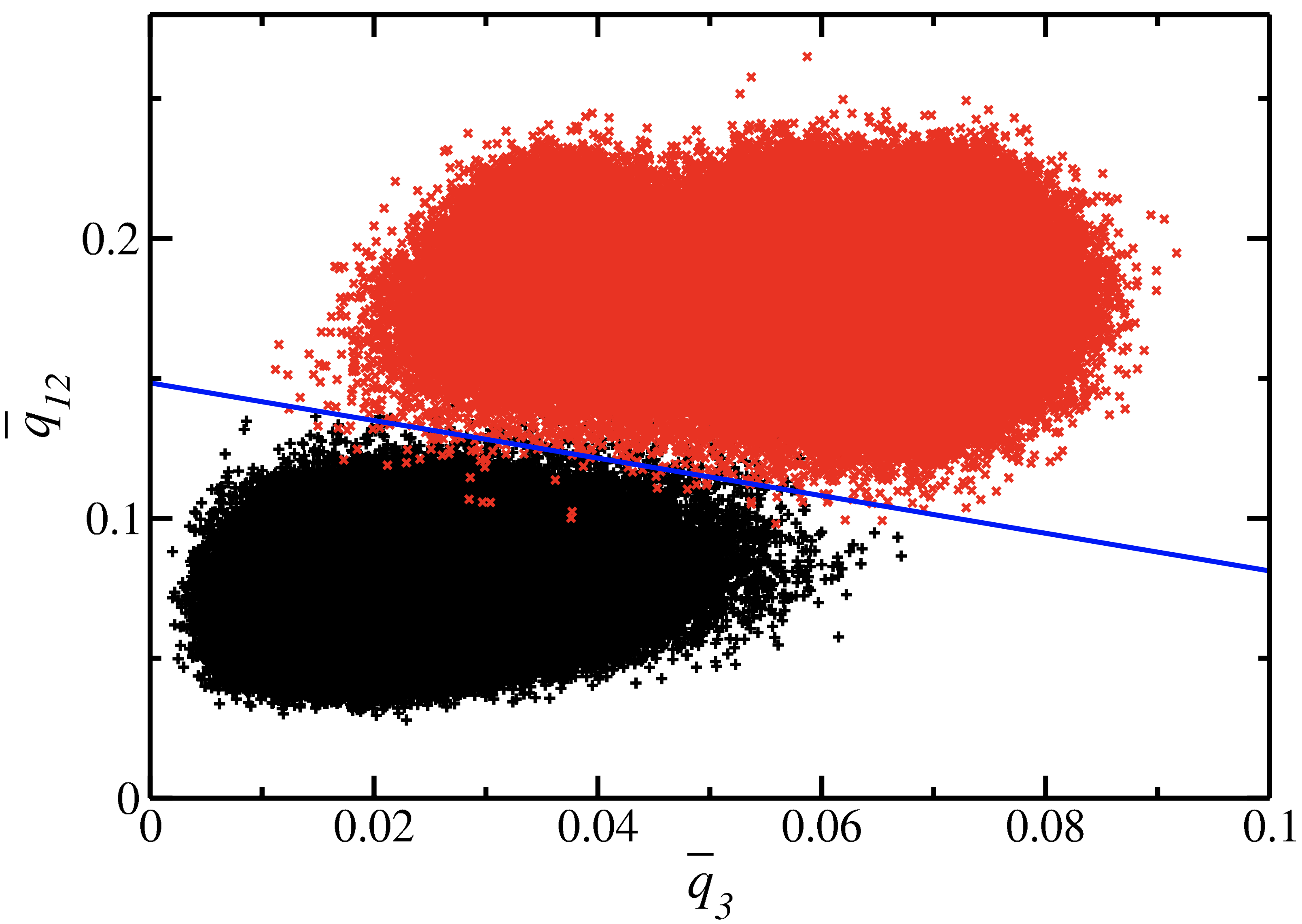}
\caption{Averaged local bond order parameters $\overline{q}_{3}$ and $\overline{q}_{12}$ of water molecules for bulk systems at $255\,\text{K}$ and $400\,\text{bar}$. Black pluses correspond to water molecules in the aqueous solution of CO$_{2}$ phase at equilibrium concentration $x^{\text{eq}}_{\text{CO}_{2}}$, red crosses represent water molecules in the hydrate phase, and the threshold with minimum mislabeling \cite{Espinosa2016c} between two phases is indicated by the blue line $\overline{q}_{c} = -0.6718\,\overline{q}_{3} + 0.1484$.}
\label{figure3}
\end{figure}

We carried out simulations in bulk phases: CO$_{2}$ hydrate and aqueous solution of CO$_{2}$ at $255\,\text{K}$  and $400\,\text{bar}$. The $\overline{q}_{3}$ and $\overline{q}_{12}$ values obtained after $50\,\text{ns}$ of production are plotted in Fig.~\ref{figure3}. As can be seen, this pair of parameters allows to differentiate clearly between the cloud of water molecules in the hydrate phase and that of water in the dissolution. 
From values plotted in Fig.~\ref{figure3} we determine a threshold function which is a linear combination of $\overline{q}_{3}$ and $\overline{q}_{12}$ parameters being $\overline{q}_{c} = -0.6718\,\overline{q}_{3} + 0.1484 $ the best separation causing a mislabeling of just $0.018 \%$. Thus this order parameter is exceptionally good at identifying solid and fluid particles. 
Finally, to determine the number of water molecules in a solid cluster we consider two molecules connected if they are labeled as solid and their separation is less than $3.5\,\text{\AA}$. The number of CO$_{2}$ is inferred by the hydrate stoichiometry 1~CO$_{2}$~:~5.75~H$_{2}$O.

It is simple to locate the transition to the solid phase using BF simulations if one has an order parameter that distinguishes reasonably well fluid and solid particles. Estimated nucleation rates do not depend on the choice of the order parameter. However, in the case of Seeding things are different. The estimate of $J$ will depend on the choice of the order parameter. Ideally one should use an order parameter that allows to estimate correctly the radius at the surface of tension of the solid cluster (see previous work for a deeper discussion of this point). In our previous work,~\cite{Grabowska2022b} we have demonstrated that the $\overline{q}_{12}$  local bond order parameters of Lechner and Dellago~\cite{Lechner2008a} is a good choice to get accurate estimates for the nucleation rates of the methane hydrate. Some of us have recently shown that the same is true when the $\overline{q}_{3}-\overline{q}_{12}$ combination is used for the methane hydrate, as well as for other hydrates, including nitrogen, hydrogen, and tetrahydrofuran hydrates.~\cite{Zeron2024b} It is necessary to show now that 
the $\overline{q}_{3}-\overline{q}_{12}$ combination is also providing good estimates of $J$ for the CO$_{2}$ hydrate within the Seeding formalism. 
The way to test that is to compare values obtained of $J$ from BF simulations to those obtained from Seeding.

\subsection{Nucleation rate by BF simulations at $255\,\text{K}$ and 
supersaturations $S=1.207$ and $S=1.268$}

 At $255\,\text{K}$ and $400\,\text{bar}$ when $S=1$ we were unable to nucleate hydrates in the 
 two phases system (CO$_{2}$ and water) within our computational 
 resources (several thousand molecules and up to 10 microseconds simulations).
 For this reason, we decided to consider two supersaturated solutions 
 at $255\,\text{K}$ and $400\,\text{bar}$, one with $S=1.207$, that corresponds to a molar fraction of $x_{\text{CO}_{2}}= 0.0969$, and another with $S = 1.268$ which corresponds to $x_{\text{CO}_{2}} = 0.1016$. Note that although both molar fractions are close to the equilibrium concentration of CO$_{2}$ in water at coexistence conditions, $x^{\text{eq}}_{\text{CO}_{2}}=0.0803$,\cite{Algaba2023a} the time required to observe nucleation in BF simulations is very different (see below).  
 The typical volume of the simulation box was 172.4 $\text{nm}^{3}$ ($S=1.207$) and $173.4\,\text{nm}^{3}$ ($S=1.268$) (containing $4942$ molecules of water and 530 or 560 molecules of CO$_{2}$ respectively). Runs were done in the isotropic $NPT$ ensemble.
 
\begin{figure}
\centering
\hspace*{-0.2cm}
\includegraphics[width=0.9\columnwidth]{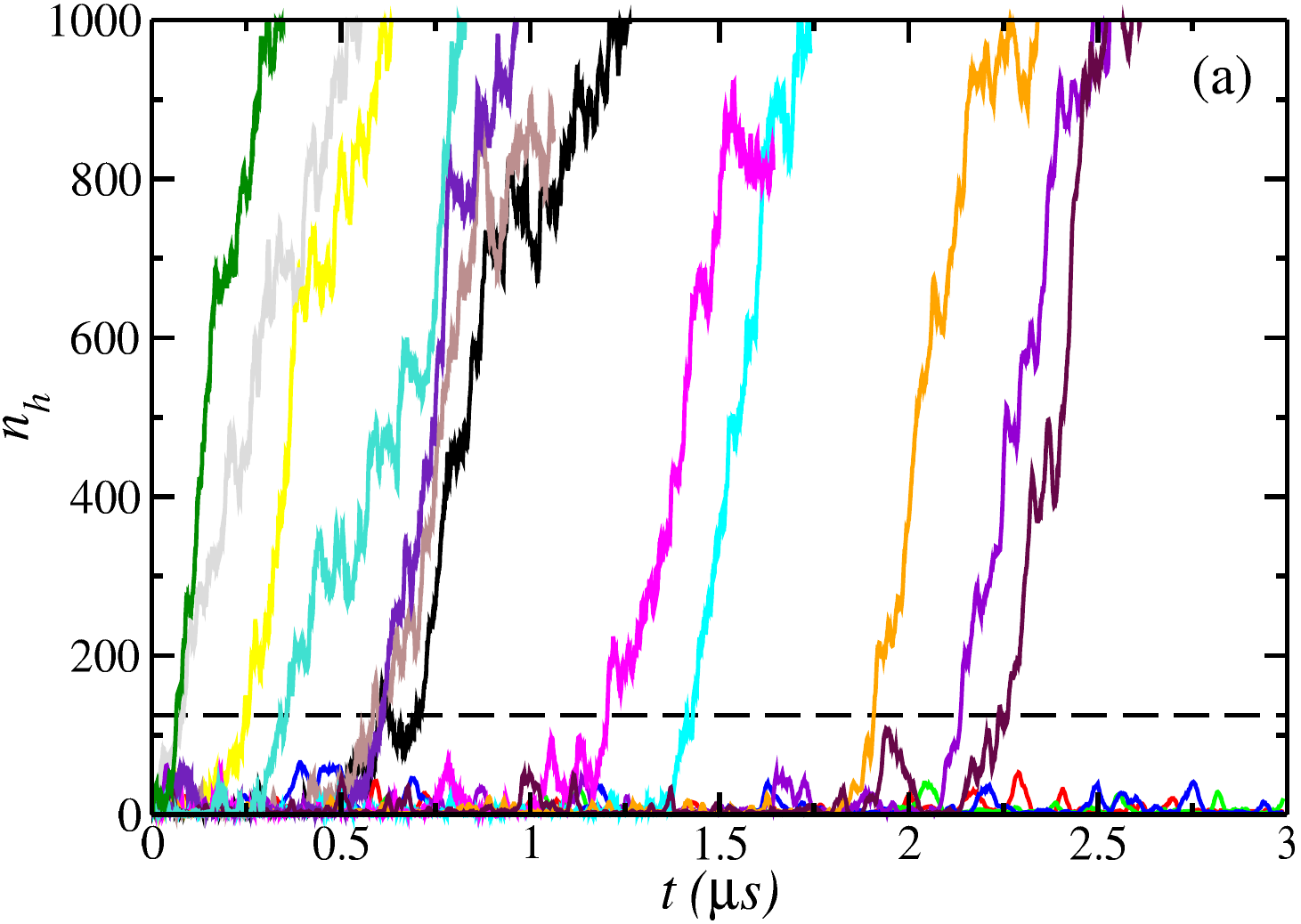}
\includegraphics[width=0.9\columnwidth]{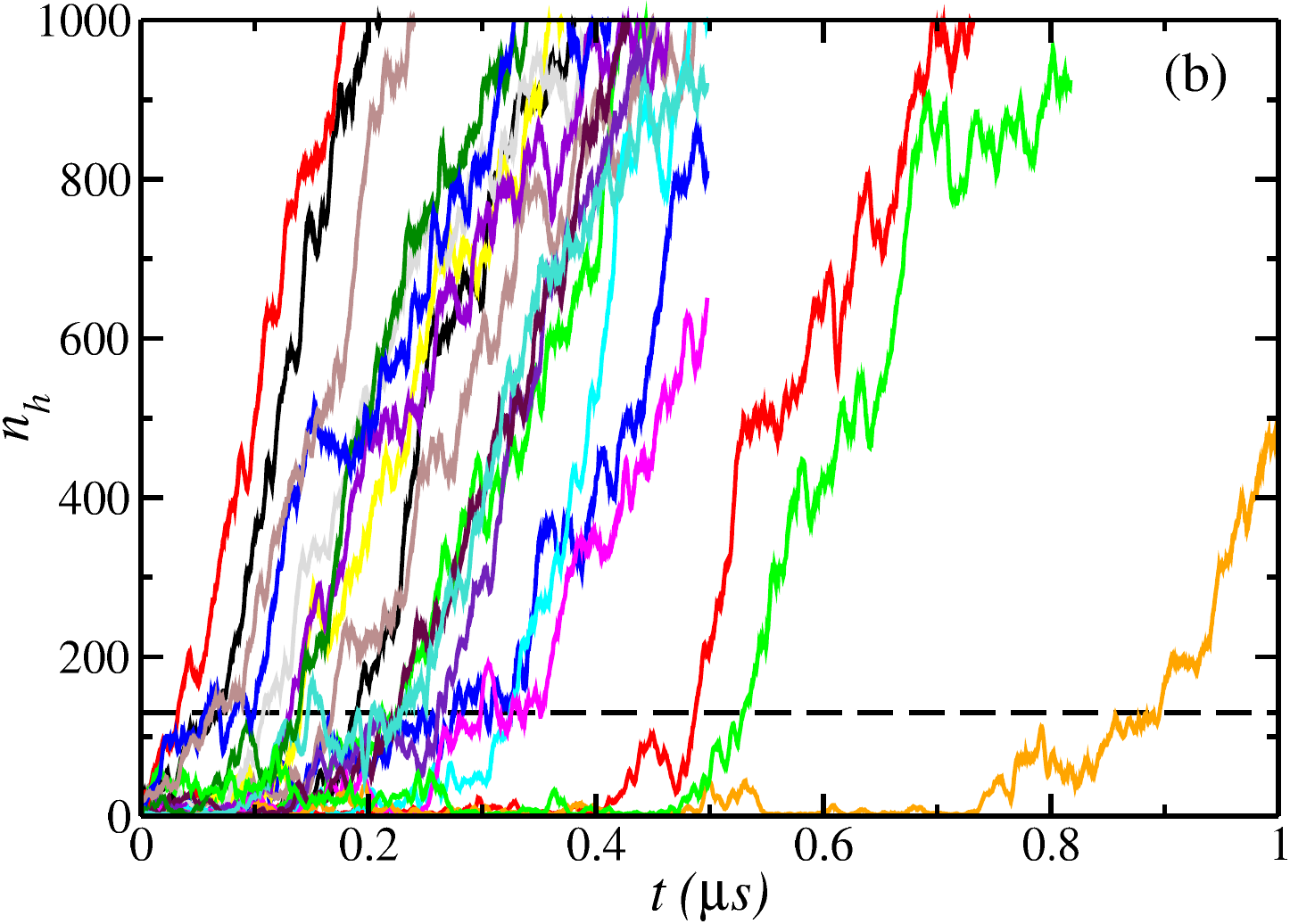}
\caption{Number of water molecules in the largest
cluster of the CO$_{2}$ hydrate, $n_{h}$, as a function of time, with supersaturation $S=1.207$ (a) and $1.268$ (b). The cluster size is obtained using the $\overline{q}_{3}-\overline{q}_{12}$ linear combination of order parameters. Each curve represents an independent BF $NPT$ simulation at $255\,\text{K}$, $400\,\text{bar}$, and the corresponding saturation. The dashed horizontal line in each panel represents a post-critical cluster that always grows irreversibly and that can be used to determine the nucleation time of each individual run.}
\label{figure5}
\end{figure}

Fig.~\ref{figure5} shows the number of water molecules in the largest solid cluster of the CO$_{2}$ hydrate, $n_{h}$, as a function of time for systems with supersaturations $S=1.207$ and $1.268$. 
In the first case ($S=1.207$), shown in panel (a), we have considered $15$ independent trajectories, and in the second case ($S=1.268$) shown in panel (b), we have simulated  20 trajectories. For each run, we determine the nucleation time as the one required to cross the horizontal line defined by $n_{h}>125$ as this corresponds to a post-critical cluster that never returns to the fluid phase and grows irreversibly. For $S=1.268$, the twenty trajectories are successful in nucleating the solid phase in less than $1\,\mu\text{s}$. For $S=1.207$, 12 out of 15 were successful after runs of up to $3\,\mu\text{s}$. 
Let us now compute the nucleation rate. For the case $S=1.268$ the nucleation rate can be estimated simply as:

 \begin{equation}
J_{\text{BF}} = \frac{1}{\tau \, V},
\label{nucleation_rate_bf}
\end{equation} 

\noindent where $\tau$  is the average time required for the system to nucleate.
For $S=1.268$ it is easy to determine this time obtaining a value of about $2\times 10^{31}/(\text{m}^3\,\text{s})$.  For $S=1.207$, not all trajectories are successful in nucleating the solid. In this case, $\tau$ could be computed from the time required to nucleate $n$ trajectories out of $n_0$ by using the expression:

\begin{equation}
\tau = \frac{ \tau_{(n_{0}-n)/n_{0}}}{\text{ln}\, \Bigl(\frac{n_{_{0}}}{n_{_{0}}-n}\Bigr)}
\label{tau_bf}
\end{equation}

Since we have performed $15$ different trajectories, $n=12$ and $n_{0}=15$, and consequently $\tau_{3/15}=2240\,\text{ns}$. Using this result, the volume of the simulation box, $V=172.4\,\text{nm}^{3}$, and combining Eqs.~\eqref{nucleation_rate_bf} and \eqref{tau_bf}, the nucleation rate of the CO$_{2}$ hydrate in the supersaturated solution with $S=1.207$ is $J_{\text{BF}}~=~4.2\,\times 10^{30}/(\text{m}^3\,\text{s})$.

Alternatively, one could follow the work of Walsh
\emph{et al.}~\cite{Walsh2011a}  and estimate $\tau$ as the total simulated time (including the full length of the run for non-successful trajectories and the time for nucleation in the successful ones and dividing by the number of successful runs which is 12 in this case). The final value using this route is $J_{\text{BF}}=3.4 \times 10^{30} / (\text{m}^{3}\,\text{s})$, which is in excellent agreement with the value obtained using the $\tau_{3/15}$ value. 

\begin{figure}
\centering
\hspace*{-0.2cm}
\includegraphics[width=0.9\columnwidth]{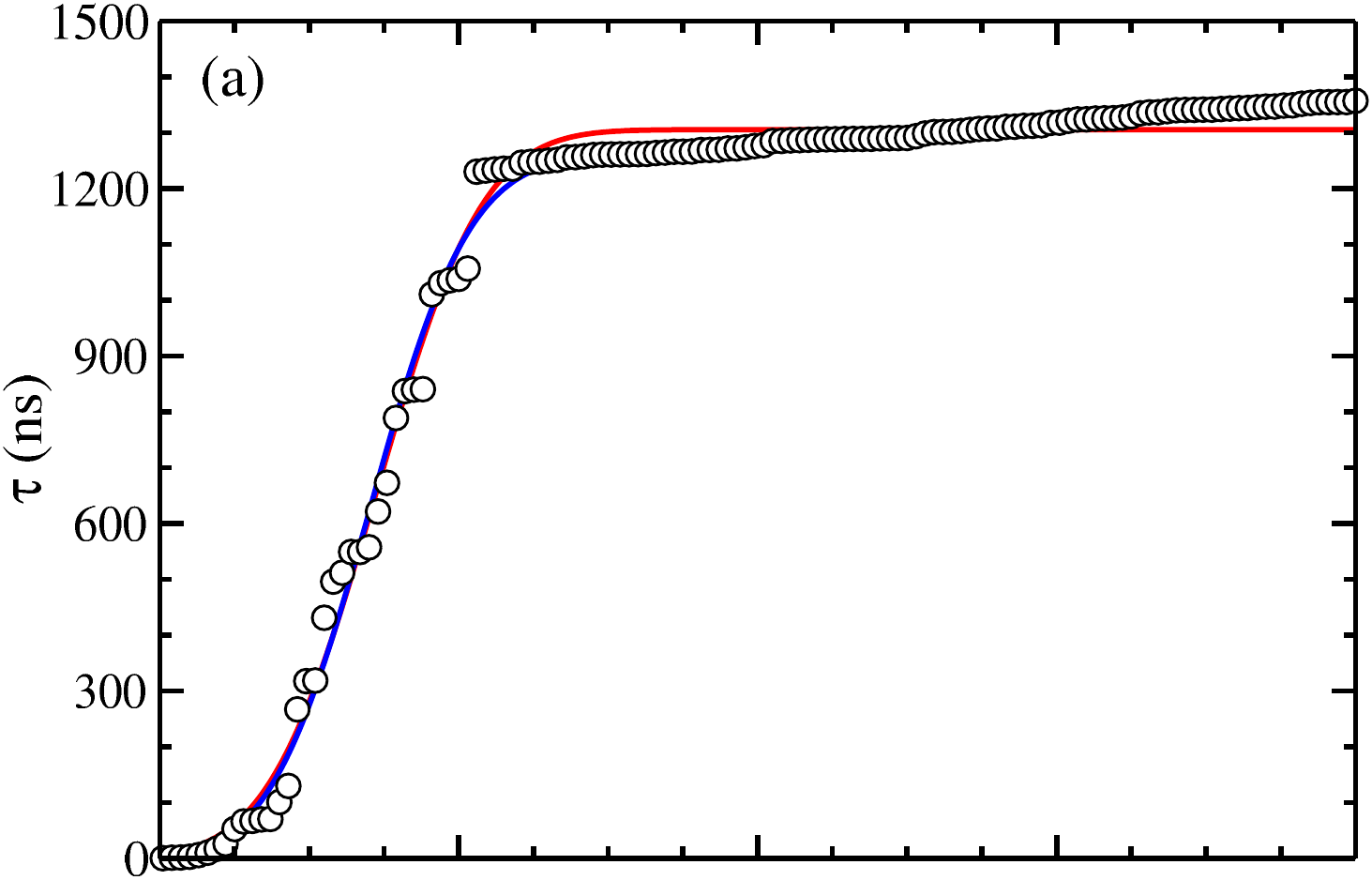}
\includegraphics[width=0.92\columnwidth]{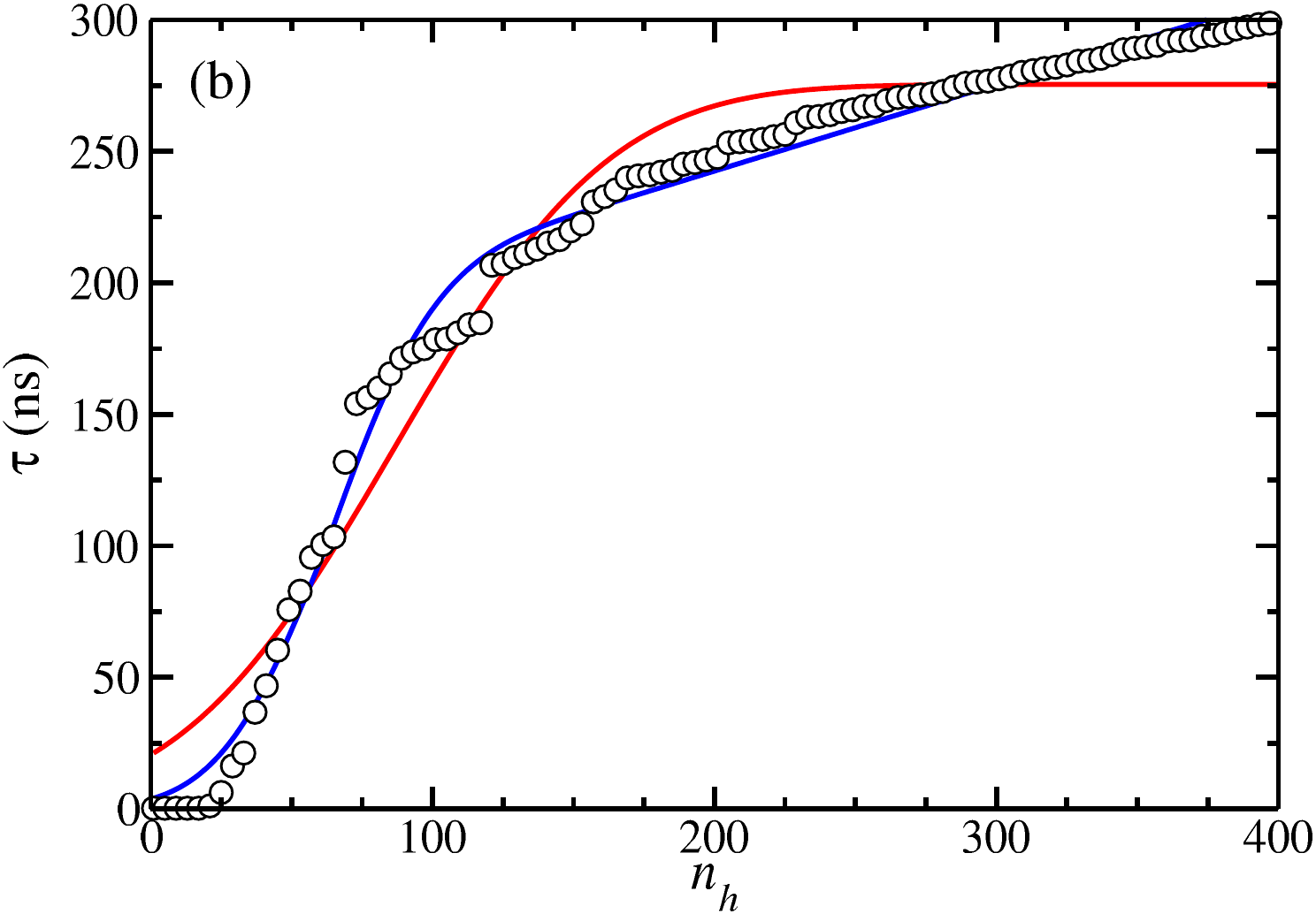}
\caption{MFPT, $\tau$, as a function of the largest cluster size, $n_{h}$, obtained for the solution of CO$_{2}$ in water at $255\,\text{K}$ and $400\,\text{bar}$  with supersaturation $S = 1.207$ (a) and $1.268$ (b). Note that $n_{h}$ is given in terms of the number of water molecules in the hydrate phase. Black circles correspond to the average time at which the cluster of water molecules in the hydrate phase reaches for the first time a certain size in the range from $0$ to $400$ molecules according to the BF simulations plotted in Fig.~\ref{figure5}. Continuous curves are fitted using Eqs.~\eqref{tau_mfpt} (red curve) and \eqref{tau_mfpt_mod} (blue curve).}
\label{FB_mfpt}
\end{figure}

A different route to determine $J$ is doing a MFPT analysis. In the MFPT analysis, $\tau(N)$, is the average elapsed time until the largest cluster of the system reaches or exceeds a threshold size $N$ for the first time. Under reasonably high barriers, $\tau(N)$ is given by the following expression,\cite{wedekind2007new,Chkonia2009a}

\begin{equation}
\tau(N) = \frac{\tau_{\text{J}}}{2}\Bigl\{1+\text{erf}\bigl[Z\sqrt{\pi}(N-N_{c})\bigr]\Bigr\}
\label{tau_mfpt}
\end{equation}

\noindent
where $\text{erf}(x)$ is the error function, $Z$ is the Zeldovich factor, $N_{c}$ critical nucleus size, and $\tau_{\text{J}}=1/J$ is the inverse of the steady-state nucleation rate $J$. This expression works well when the growth's time scale is small compared with the time scale for nucleation. 
Alternatively when they are comparable one could fit the results into the expression:

\begin{equation}
    \tau_{mod}(N) = \tau(N) + \frac{1}{2G} (N - N_{c})\Bigl\{1+\text{erf}\bigl[C(N - N_{c})\bigr]\Bigr\},
    \label{tau_mfpt_mod}
\end{equation}
where $G$ is the growth rate and  $C$ is a positive constant and $\tau(N)$ is given by Eq.~\eqref{tau_mfpt}. 
In Fig.~\ref{FB_mfpt}, a MFPT analysis is performed and the results are fitted to both Eqs.~\eqref{tau_mfpt} (red lines) and \eqref{tau_mfpt_mod} (blue lines). The value of $J$ is obtained from the MFPT analysis as:

\begin{equation}
    J_{\text{MFPT}} = \frac{1}{\tau_{j} \, V}
    \label{nucleation_rate_mfpt}
\end{equation}

\noindent The results for the nucleation rate obtained from the MFPT analysis are shown in Table \ref{table_mfpt}. Note that the results of Fig.~\ref{FB_mfpt} show that for the two supersaturations studied ($S=1.207$ and $1.268$), the solid is formed by the nucleation of just one critical cluster (after an induction time) followed by growth. However, for much higher supersaturations, one would expect the appearance of multiple small critical clusters so that the solid could grow via the growth of these individual clusters~\cite{Chkonia2009a} and by the Ostwald ripening mechanism.~\cite{Weijs2012a,Xu2025a}

 \begin{table}
    \centering
    \caption{Nucleation rate of CO$_{2}$ hydrate in water, $J$, at $255\,\text{K}$, $400\,\text{bar}$, and supersaturation S = 1.207 and S = 1.268 using the MFPT method.}
    \begin{tabular}{lcccccccc}
    \hline
    \hline
        \multirow{2}{4em}{$S$} & \,\, & \multicolumn{3}{c}{1.207} & \,\, & \multicolumn{3}{c}{1.268} \\
    \cline{3-5}
    \cline{7-9}
         & \,\, & Eq.~\eqref{tau_mfpt} & \,\, & Eq.~ \eqref{tau_mfpt_mod} & \,\, & Eq.~\eqref{tau_mfpt} & \,\, & Eq.~\eqref{tau_mfpt_mod} \\
    \hline
        $\tau_{j}\,\text{(ns)}$ & \,\, & 1305.6  & \,\, & 1232.4 & \,\, & 275.5 & \,\, & 197 \\
        $Z$              & \,\, & 0.014   & \,\, & 0.015   & \,\, & 0.007 & \,\, & 0.013 \\
        $N^{\text{H}_{2}\text{O}}_{c}$ & \,\, & 72.2    & \,\, & 69.8    & \,\, & 86.7 & \,\, & 61.7 \\
        $G\,(\text{ns}^{-1})$       & \,\, & - -     & \,\, & 2.64     & \,\, & - - & \,\, & 3.03\\
        $C$              & \,\, & - -     & \,\, & 0.19     & \,\, & - - & \,\,  & 179.7 \\
        $V\,(\text{nm}^{3})$   & \,\, & 172.4   & \,\, & 172.4   & \,\, & 173.4 & \,\,  & 173.4 \\
        $J\,(\text{m}^{-3}\,\text{s}^{-1})$ & \,\, & 4.4 $\times 10^{30}$ & \,\, &  $4.7 \times 10^{30}$ & \,\, & 2.1 $\times 10^{31}$ & \,\, & 2.9 $\times 10^{31}$ \\
    \hline
    \hline        
    \end{tabular}
    \label{table_mfpt}
\end{table}

The summary is that BF simulations lead to values of $J$  of about $4\times 10^{30}$  and $2\times 10^{31}/\text{(m}^3\,\text{s)}$ for $S=1.207$ and $S=1.268$, respectively. 
For methane hydrate one obtained similar values of $J$ for 
$S=4.72$ and $5.67$, respectively. Thus nucleation of CO$_{2}$ hydrate is easier since it appears at lower supersaturations. What provokes this enhancement of homogeneous nucleation in the CO${_2}$ hydrate?
Certainly CO${_2}$ is about one order of magnitude more soluble than CH${_4}$ at the same pressure and 
supercooling (i.e.,  $x_{\text{CO}_{2}}=0.0803$ for CO${_2}$ versus  $x_{\text{CH}_{4}}=0.0089$ for methane). 
However CH${_4}$ seems more efficient. In fact, it is able to reach values of $J$ of the order of $10^{30}$ with a concentration of  $x_{\text{CH}_{4}}=0.042$ whereas for CO${_2}$ one needs a concentration of $x_{\text{CO}_{2}}=0.097$ to obtain the same nucleation rate (a similar conclusion was obtained in previous work by some of us on the growth rate of the hydrate\cite{blazquez2023growth}). Later in this paper, we will try to identify the key ingredient that makes the homogeneous nucleation of the CO${_2}$ hydrate much easier.

The values of $J$ for $S=1.207$ of this section will allow to determine if the choice of order parameters to distinguish between hydrate-like and liquid-like water molecules can be used with confidence to correctly determine nucleation rates when using the Seeding technique. Note that $J$ values using this technique are quite sensitive to the choice of the order parameter in contrast with BF runs, which do not depend much on the choice of the order parameter.

\subsection{Nucleation rate from Seeding simulations at $T=255\,\text{K}$ and supersaturation $S=1.207$}

\begin{figure}
\centering
\hspace*{-0.2cm}
\includegraphics[width=0.9\columnwidth]{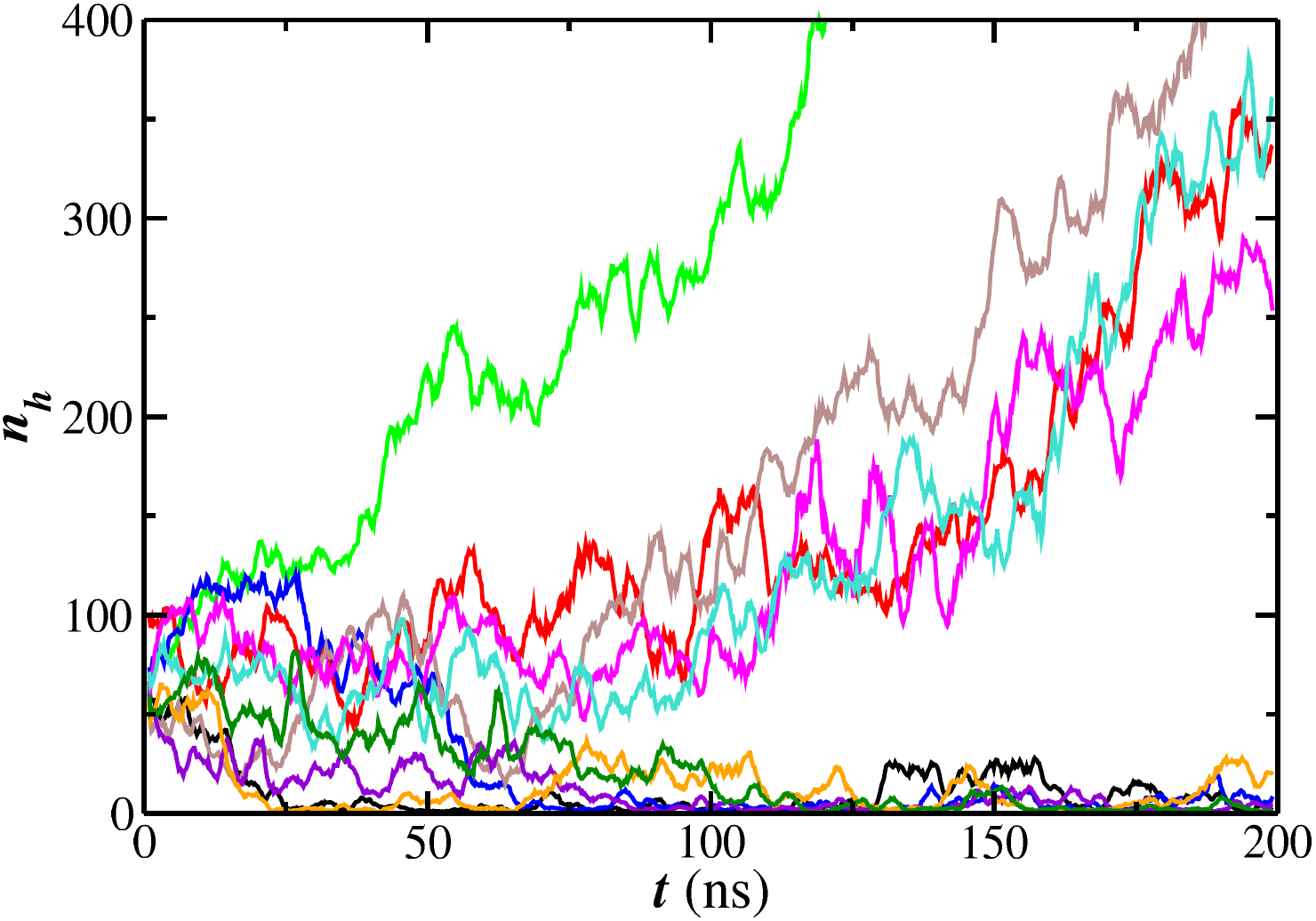}
\caption{Number of water molecules in the largest
cluster of the CO$_{2}$ hydrate, $n_{h}$, as a function of time, for a supersaturated solution of CO$_{2}$ in water ($S = 1.207$) at $255\,\text{K}$ and $400\,\text{bar}$. The starting configuration contains a seed of hydrate of radius $r=0.79\, \text{nm}$, which is critical at these conditions. The average size of the cluster, $N_{c}^{\text{H}_{2}\text{O}} = 55$, is obtained using the $\overline{q}_{3}-\overline{q}_{12}$ linear combination shown in Fig. \ref{figure3}.}
\label{nh_seeding_SS}
\end{figure}

The Seeding method was implemented as follows. 
After equilibrating a one-phase system using isotropic $NPT$ simulations at $255\,\text{K}$ and $400\,\text{bar}$ with $S=1.207$, we inserted spherical hydrate seeds of different sizes as it is schematized in Fig.~\ref{figure2}b.
After insertion, we removed particles that overlap with the solid cluster and allowed for a short run where the seed molecules were frozen. 
After that several $NPT$ runs (with all molecules free to move) with different initial random velocities were performed. When the seed was small the solid cluster quickly melted. When the seed was large the solid cluster grew. Just at the critical size, there is a probability of $50\%$ that the hydrate grows or melts. We considered $9$ different cluster sizes:  $r=0.51$, $0.61$, $0.68$, $0.74$, $0.79$, $0.85$, $0.87$, $0.91$, and $0.95\,\text{nm}$, each of them formed from $15$, $25$, $35$, $45$, $55$, $65$, $75$, $85$, and $95$ water molecules in average. For each cluster size, we have performed $10$ different simulations. We have observed that for spherical hydrate seeds with a radius lower than $0.74\,\text{nm}$ only $2$ or $3$ trajectories grow ($2$ of $10$ for the two lowest radii and $3$ of $10$ for $r=0.68\,\text{nm}$. On the contrary, for  
spherical hydrate seeds with a radius equal or greater than $0.86\,\text{nm}$ most of the trajectories grow ($6$ of $10$ for $r=0.85\,\text{nm}$ and $9$ of $10$ for $r\ge 0.87\,\text{nm}$). According to this, the spherical hydrate seed that can be considered critical is that with $r=0.79\,\text{nm}$, formed from $55$ water molecules. As can be seen in Fig.~\ref{nh_seeding_SS}, when a spherical hydrate seed of radius $r=0.79\,\text{nm}$ is inserted into the supersaturated solution $S = 1.207$, at $255\,\text{K}$ and $400\,\text{bar}$, the system behaves as critical showing 5 trajectories for which the inserted seed grows and $5$ in which rapidly melts.
The initial size of the seed is calculated by averaging the largest cluster size during the equilibration period of $2\,\text{ns}$ in all runs using the selected parameters. 

\begin{figure}
\centering
\hspace*{-0.2cm}
\includegraphics[width=0.9\columnwidth]{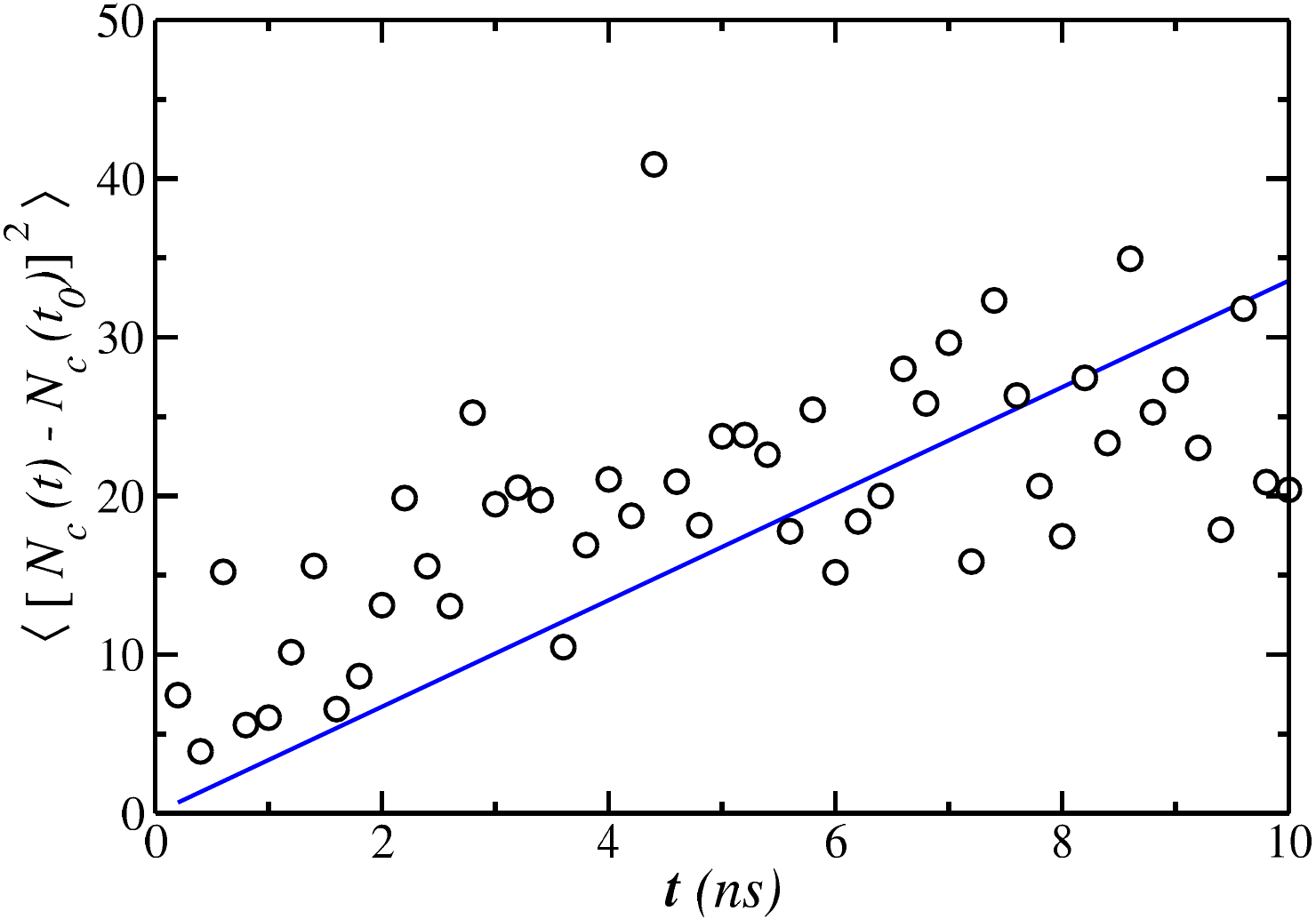}
\caption{$\langle [N_{c}(t) - N_{c}(t_{0})]^{2} \rangle$ factor, given in terms of CO$_{2}$ molecules, averaged over $10$ independent simulations of a supersaturated solution of CO$_{2}$ in water ($S = 1.207$) with the critical seed of hydrate at $255\,\text{K}$ and $400\,\text{bar}$ plotted in
Fig.~\ref{nh_seeding_SS}. Black circles represent values obtained from simulations and the blue line represents the linear fit of the simulation results.}
\label{attachment_r131_S121}
\end{figure}

Once we know the critical cluster size, the attachment rate $f^{+}_{\text{CO}_{2}}$ can be calculated by averaging the squared difference between the initial cluster size and the cluster size in time. This term behaves linearly and $f^{+}_{\text{\text{CO}}_{2}}$ is defined as half of the slope of the linear fit according to Eq.~\eqref{attachment_rate}.
Applying this formula to all Seeding runs, we obtain the behavior plotted in Fig.~\ref{attachment_r131_S121} and $f^{+}_{\text{CO}_{2}}= 1.68 \times 10^{9} \, /\text{(s)}$.
Besides, from our previous work, the driving force at these thermodynamic conditions is $\Delta \mu_{\text{N}} = -2.73 \,k_{B}T$. In this way, the Zeldovich factor, Eq.~\eqref{zeldovich} is $Z=0.123$ and using Eq.~\eqref{nucleation_rate_cnt_2} we have estimated the nucleation rate
$J= 1.4 \times 10^{30} / \text{(m}^{3}\,\text{s)}$ for a supersaturated solution $S = 1.207$ at $255\,\text{K}$ and $400\,\text{bar}$ via Seeding approach.
As can be noticed, this result is in complete agreement with the findings using BF simulations. According to this, the linear combination of $\overline{q}_{3}$ and $\overline{q}_{12}$ can be safely used to describe the correct cluster size. Results of this section are summarized in Table \ref{table_seeding}.

\subsection{Seeding simulations of BF clusters at $T=255\,\text{K}$ and supersaturation $S=1.207$}

The formation of hydrates from solutions with appropriate composition of the guest component using BF simulations exhibits multiple pathways, including amorphous agglomeration of cages, partially-ordered hydrates, and mixtures of different crystal structures among others, as clearly explained by Guo, Zhang and collaborators.~\cite{Guo2011a,Zhang2015a} The nuclei formed during BF simulations may not exhibit the thermodynamically stable sI crystallographic structure, although as Zhang \emph{et al.}~\cite{Zhang2015a} have shown, it is possible to get spontaneously formed cluster with a high degree of sI crystallinity. Jacobson and Molinero have also analyzed the role of amorphous intermediates in the formation of clathrate hydrates.~\cite{Jacobson2011a}

\begin{figure}
\centering
\hspace*{-0.2cm}
\includegraphics[width=0.9\columnwidth]{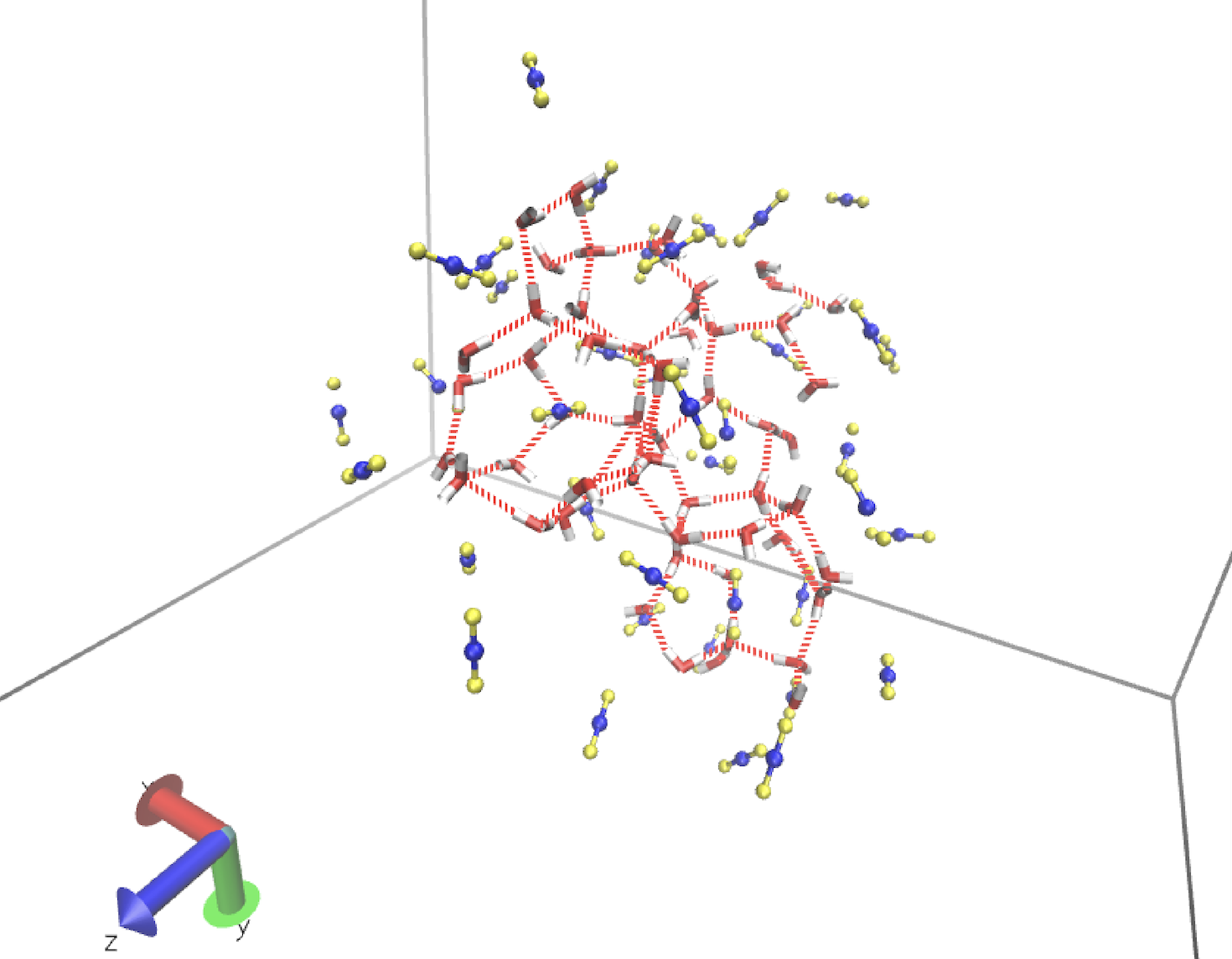}
\caption{Snapshot of cages of the CO$_{2}$ hydrate taken from a BF simulation, with $S=1.207$, at $255\,\text{K}$ and $400\,\text{bar}$, forming a cluster with $55$ water molecules. This cluster has been extracted from one of the BF trajectories shown in Fig.~\ref{figure5}a. Water molecules are
represented using red sticks for the oxygen atoms and white ticks for the hydrogen atoms. Red dashed lines represent hydrogen bonds between the molecules of water in the cluster, and CO$_{2}$ molecules are represented using blue sticks for the carbon atoms and red
yellow for the oxygen atoms.} 
\label{seed-55wat-co2}
\end{figure}

\begin{figure}
\centering
\hspace*{-0.2cm}
\includegraphics[width=0.9\columnwidth]{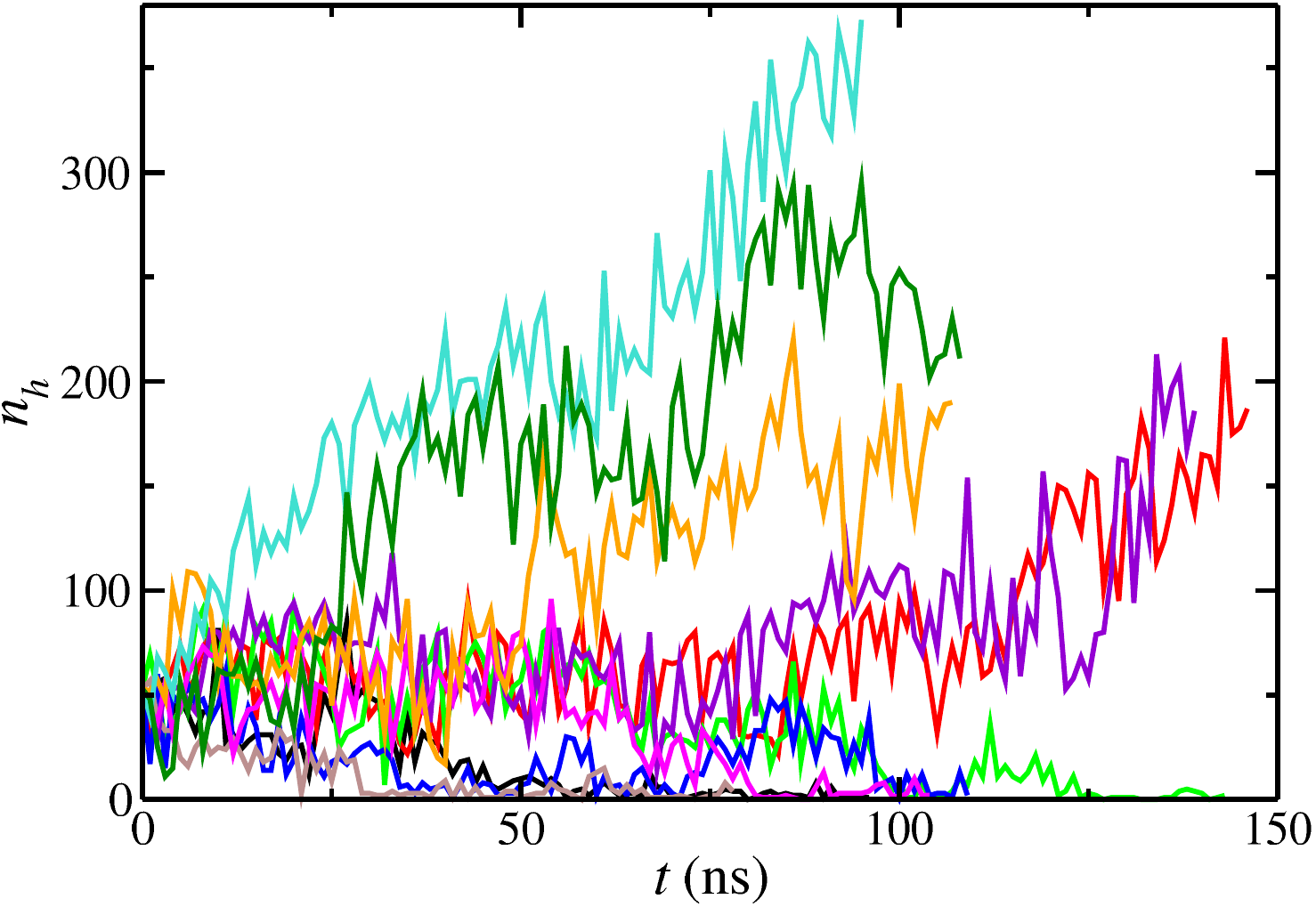}
\caption{Number of water molecules in the largest
cluster of the CO$_{2}$ hydrate, $n_{h}$, as a function of time, for a supersaturated solution of CO$_{2}$ in water ($S = 1.207$) at $255\,\text{K}$ and $400\,\text{bar}$. The starting configuration contains a seed of hydrate from BF simulations at the same thermodynamics conditions. The starting size of the cluster, $N_{c}^{\text{H}_{2}\text{O}} = 55$, is obtained using the $\overline{q}_{3}-\overline{q}_{12}$ linear combination shown in Fig. \ref{figure3}.}
\label{nh_S121_nonsphe-53wt}
\end{figure}

In Section III.B we have obtained estimations of the CO$_{2}$ hydrate nucleation rate at $255\,\text{K}$ and $400\,\text{bar}$, with supersaturation $S=1.207$, from BF simulations. The value reported there is $J_{BF}\sim 10^{30}\,\text{m}^{-3}\text{s}^{-1}$. We have also used the Seeding Technique to estimate the nucleation rate of the hydrate at the same thermodynamic conditions and supersaturation (Section III.C). The value obtained is of the same order of magnitude, $J\sim10^{30}\,\text{m}^{-3}\text{s}^{-1}$. It is possible to analyze the clusters used in BF and Seeding simulations to obtain additional information from these two embryos. Particularly, one could use a nucleus generated from BF simulations as a seed in Seeding simulations, i.e., to insert a nucleus formed during BF simulations. This allows us to check if two hydrate clusters formed from the same number of molecules, one obtained from BF simulations and a perfect (sI) and spherical one usually used in Seeding simulations, are critical. Following this approach, we have randomly selected a trajectory of our BF simulations with $S=1.207$ (one of those shown in Fig.~\ref{figure5}) and picked up a solid hydrate cluster formed from $55$ water molecules from the corresponding trajectory. Fig.~\ref{seed-55wat-co2} shows a snapshot of this cluster that has the same number of water molecules as the critical one used in the Seeding simulations (see Table~II). We insert the cluster obtained from BF simulations in the aqueous solution as it was done in Section III.C and run $10$ different independent trajectories. If the BF cluster is critical, the system should show $5$ trajectories for which the inserted seed grows and $5$ in which it rapidly melts. Fig.~~\ref{nh_S121_nonsphe-53wt} shows the number of water molecules of this CO$_{2}$ hydrate cluster, $n_{h}$, as a function of time, in the supersaturated solution of CO$_{2}$ in water ($S = 1.207$) at $255\,\text{K}$ and $400\,\text{bar}$. As can be seen, our results indicate that the cluster obtained from BF simulations, with the same size as a cluster that is critical according to Seeding simulations, is also critical (at the studied conditions). Notice that Guo and Zhang~\cite{Guo2021a} found smaller sizes of the critical cluster when amorphous clusters were considered when compared to crystalline ones. This is an interesting observation that deserves to be analyzed in more detail in the future. However, at least for the case considered here ($S=1.207$) we found that the size of a crystalline critical cluster and a critical cluster obtained from BF simulations is rather similar.

\subsection{Nucleation rate by Seeding simulations at $T=255\,\text{K}$ and $S=1$}
\label{sec:seeding}

We were not able to nucleate the hydrate in BF runs at $255\,\text{K}$ and $400\,\text{bar}$ when having the two-phase system with  CO${_2}$ and water at equilibrium (i.e., $S=1$). Thus we shall use the Seeding method to estimate the nucleation rate after having validated the technique with the results of the previous subsection. 
The Seeding method was implemented as follows. 
We first constructed the starting configuration, as shown in Fig.~\ref{figure2}c, by equilibrating in the isotropic $NPT$ ensemble a cubic simulation box formed from $12000$ water molecules and $1048$ CO$_{2}$, i.e., $x_{\text{CO}_2} = 0.0803$ ($S=1$). Once the temperature, pressure, and average volume achieved a constant value, we add a reservoir of liquid CO$_{2}$ at both sides of the previous dissolution forming two planar interfaces with $4952$ CO$_{2}$ molecules in total, including the reservoir and solution. The $z$-axis direction is perpendicular to the CO$_{2}$-water interface. Again, this two-phase system is equilibrated in an $NP_{z}\mathcal{A}T$ ensemble keeping constant the cross-section area, $\mathcal{A}$, with the value being the average area found in the equilibration part before putting the reservoir.
We now inserted spherical seeds of CO$_{2}$ hydrate of radius between $1.0$ and $1.5\,\text{nm}$ in the middle of the aqueous phase, removed overlapping particles in the solution, and equilibrated for one or $2\,\text{ns}$. We then performed $NP_z\mathcal{A}T$ runs. The length of these runs was about $200\,\text{ns}$.  
The size of the system (although it fluctuates in the $z$ direction) is about  $7.4 \times 7.4 \times 12.4$\,nm$^{3}$.

\begin{figure}
\hspace*{-0.2cm}
\includegraphics[width=0.9\columnwidth]{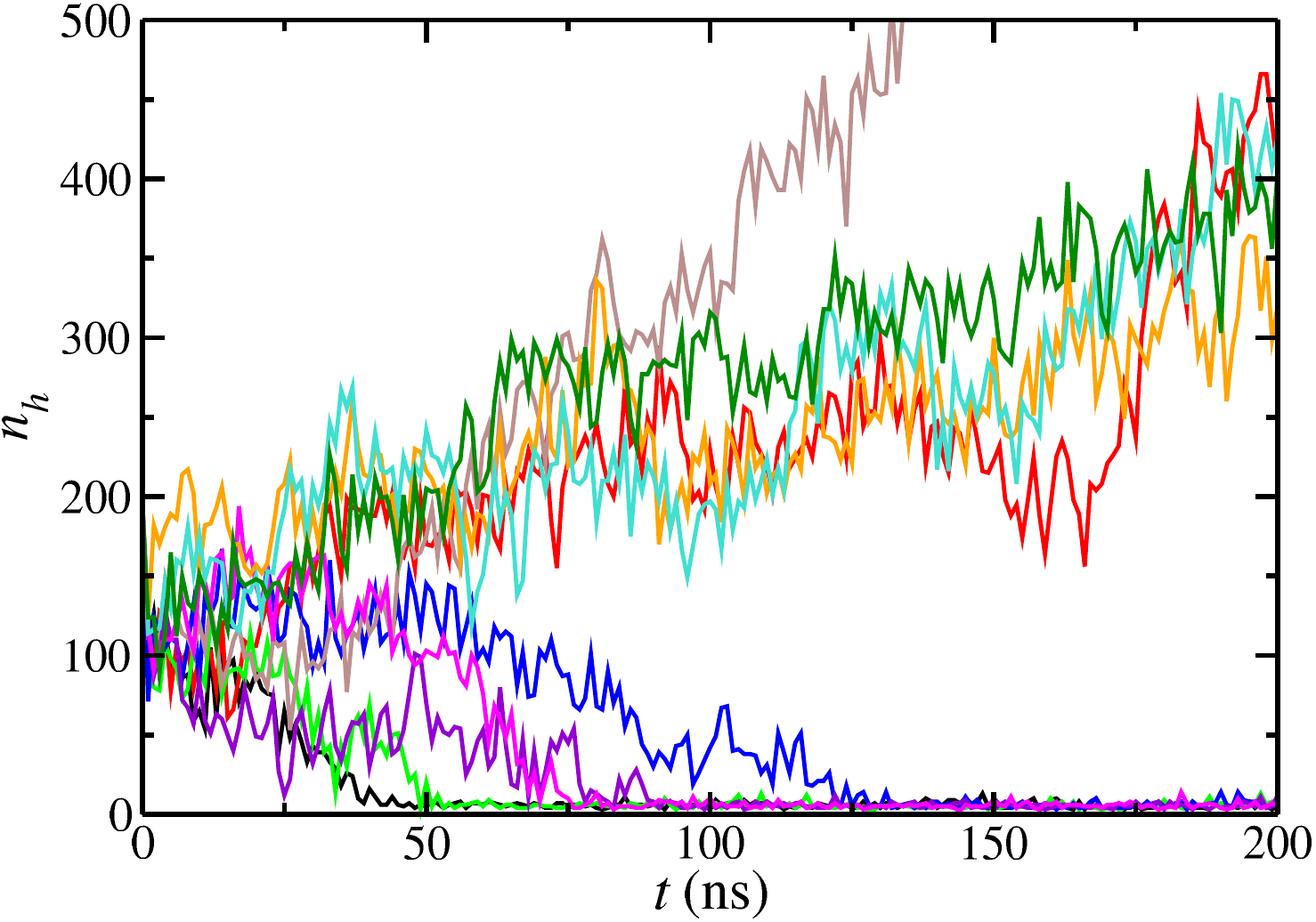}
\caption{Number of water molecules in the largest
cluster of the CO$_{2}$ hydrate, $n_{h}$, as a function of time, for the saturated solution of CO$_{2}$ in water ($S=1$) at $255\,\text{K}$ and $400\,\text{bar}$. The starting configuration contains a seed of hydrate of radius $r=1.01\, \text{nm}$, which is critical at these conditions. The average size of the cluster, $N_{c}^{\text{H}_{2}\text{O}} =115$, is obtained using the $\overline{q}_{3}-\overline{q}_{12}$ linear combination of the local bond order parameters.}
\label{nh_seeding_S1}
\end{figure}

\begin{figure}
\centering
\hspace*{-0.2cm}
\includegraphics[width=0.9\columnwidth]{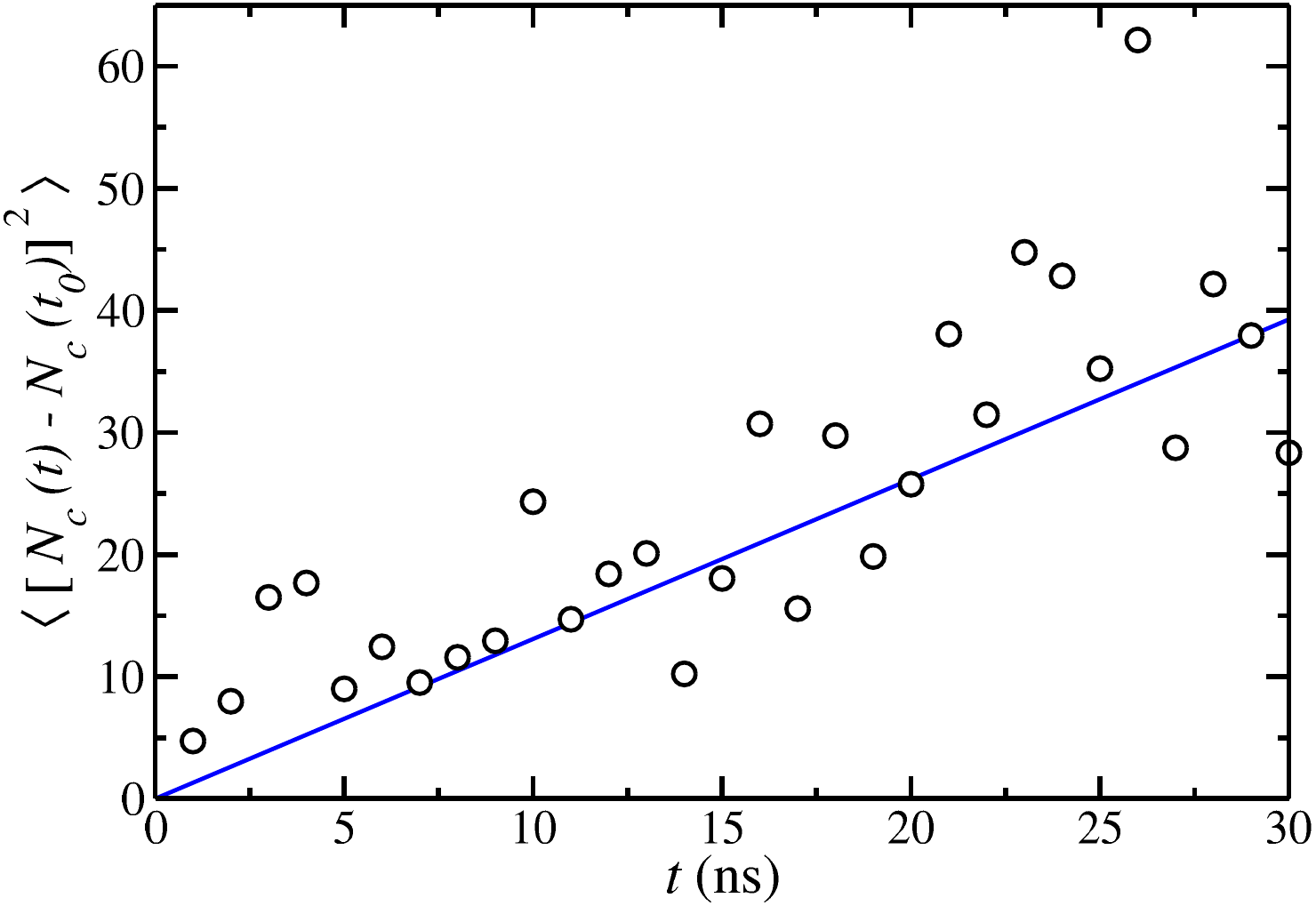}
\caption{$\langle [N_{c}(t) - N_{c}(t_{0})]^{2} \rangle$ factor, given in terms of CO$_{2}$ molecules, averaged over $10$ independent simulations of the saturated solution of CO$_{2}$ in water ($S=1$) with the critical seed of hydrate at $255\,\text{K}$ and $400\,\text{bar}$ plotted in
Fig.~\ref{nh_seeding_S1}. Black circles represent values obtained from simulations and the blue line represents the linear fit of the simulation results.}
\label{attachment_r144_S1}
\end{figure}

The size of the largest cluster, as a function of time, is plotted in Fig.~\ref{nh_seeding_S1} for an initial seed of radius $r=1.01\,\text{nm}$.  As can be seen, when the size of the largest cluster is about 115(5) water molecules the cluster becomes critical and thus the seed  melts in half of the trajectories and grows in the other half. Notice that this number of water molecules in the hydrate phase corresponds to 19(1) CO$_{2}$ molecules also in this phase. The attachment rate can be calculated through the linear fit of $\langle [ N_{c}^{\text{CO}_{2}} (t) - N_{c}^{\text{CO}_{2}} (t_{0})]^{2} \rangle$, as a function of time, at this condition as is shown in Fig. \ref{attachment_r144_S1}. In this case we estimate $f^{+}_{\text{CO}_{2}} = 6.54 \times 10^{8}\,\text{s}^{-1}$.
Using Eq.~\eqref{driving_force_cnt}, we find that the free energy barrier of nucleation for the system of CO$_{2}$ in water at $255\,\text{K}$, $400\,\text{bar}$, and concentration $S=1$ is $\Delta G_{c}=22(2)\,k_{B}T$, which is about 5 times less than that in the case of CH$_{4}$ in water at the same supercooling ($\Delta G_{c} =95\,k_{B}T$ as we found in our previous work~\cite{Grabowska2022b}). The Zeldovich factor is thus $Z=0.077$ and the nucleation rate estimated using the linear combination of the $\overline{q}_{3}$ and $\overline{q}_{12}$ order parameters and Eq.~\eqref{nucleation_rate_cnt_2} is $J = 2(5) \times 10^{25}\,\text{m}^{-3}\text{s}^{-1}$.
All results required to estimate the nucleation rate from Seeding are shown in Table \ref{table_seeding}. Our estimate of $J$ at $255\,\text{K}$ and $400\,\text{bar}$, for $S=1$ (i.e., under experimental conditions), namely, $J = 2(5) \times 10^{25}\,\,\text{m}^{-3}\text{s}^{-1}$ is consistent with the value reported at $260\,\text{K}$ and $500\,\text{bar}$ by Arjun and Bolhuis,\cite{Arjun2021a} $J = 1 \times 10^{26}\,\,\text{m}^{-3}\text{s}^{-1}$. However, it should be noticed that: (1) the force field used here is similar but not identical to that used by Arjun and Bolhuis (we include here deviations from the Lorentz-Berthelot energetic combining rule for the interaction between the carbon atom of CO$_2$ and the oxygen of water in contrast to Arjun and Bolhuis); (2) the thermodynamic conditions are slightly different; and (3) Arjun and Bolhuis used a bubble of CO$_2$ as a reservoir and therefore the solubility of the gas was higher than that of the planar interface implemented in this work. In any case, even taking these differences into account, it seems that the results of this work are consistent with those of Arjun and Bolhuis.~\cite{Arjun2021a} 
 
The homogeneous nucleation rate at experimental conditions for $400\,\text{bar}$ and $35\,\text{K}$ of supercooling is huge. In fact, it is about $30$ orders of magnitude larger than that found for methane under the same conditions (it was found of the order of $10^{-7}\,\,\text{m}^{-3}\text{s}^{-1}$). Note that the comparison is performed at the same pressure ($400\,\text{bar}$) and supercooling ($\Delta T=35\,\text{K}$). Therefore, homogeneous nucleation is significantly more important in CO${_2}$ than in CH$_{4}$ and will be present in experiments at much higher temperatures. That leads to a very interesting question: 
what is the factor provoking such a huge difference of $J$ value? In this context, it is relevant to mention the work of Zhang \emph{et al.}~\cite{Zhang2018b} These authors proposed a novel explanation for the dependence of the self-diffusion coefficient of guest molecules on guest concentration. They suggested that the higher mobility of CO$_{2}$ in water, compared to CH$_{4}$, necessitates a greater concentration of CO$_{2}$ in water (relative to methane) to induce nucleation. In other words, they established a connection between guest dynamics and hydrate nucleation. However, as we demonstrate in Section III.E, although there could be a contribution of the CO$_{2}$ mobility, we believe that the primary factor behind the 30-order-of-magnitude difference in nucleation rates is the disparity in interfacial free energy between the hydrate and aqueous solution for each hydrate. Interestingly, the mobility only enters in the attachment rate, which exhibits similar values in both hydrates in the conditions considered in this work ($f^{+}_{\text{CO}_{2}}\approx 6\times10^{8}\,\text{s}^{-1}$ and $f^{+}_{\text{CH}_{4}}\approx 1\times 10^{9}\,\text{s}^{-1}$). However, we think the main reason for the huge difference between the $J$ values of the CO$_{2}$ and CH$_{4}$ hydrates is due to the difference of the nucleation barriers of both hydrates, $\Delta G_{c}\sim22\,k_{B}T$ and $\sim95\,k_{B}T$, for the CO$_{2}$ and CH$_{4}$ hydrate, respectively, as we discuss in Section III.F.

\begin{figure}
\centering
\hspace*{-0.2cm}
\includegraphics[width=0.9\columnwidth]{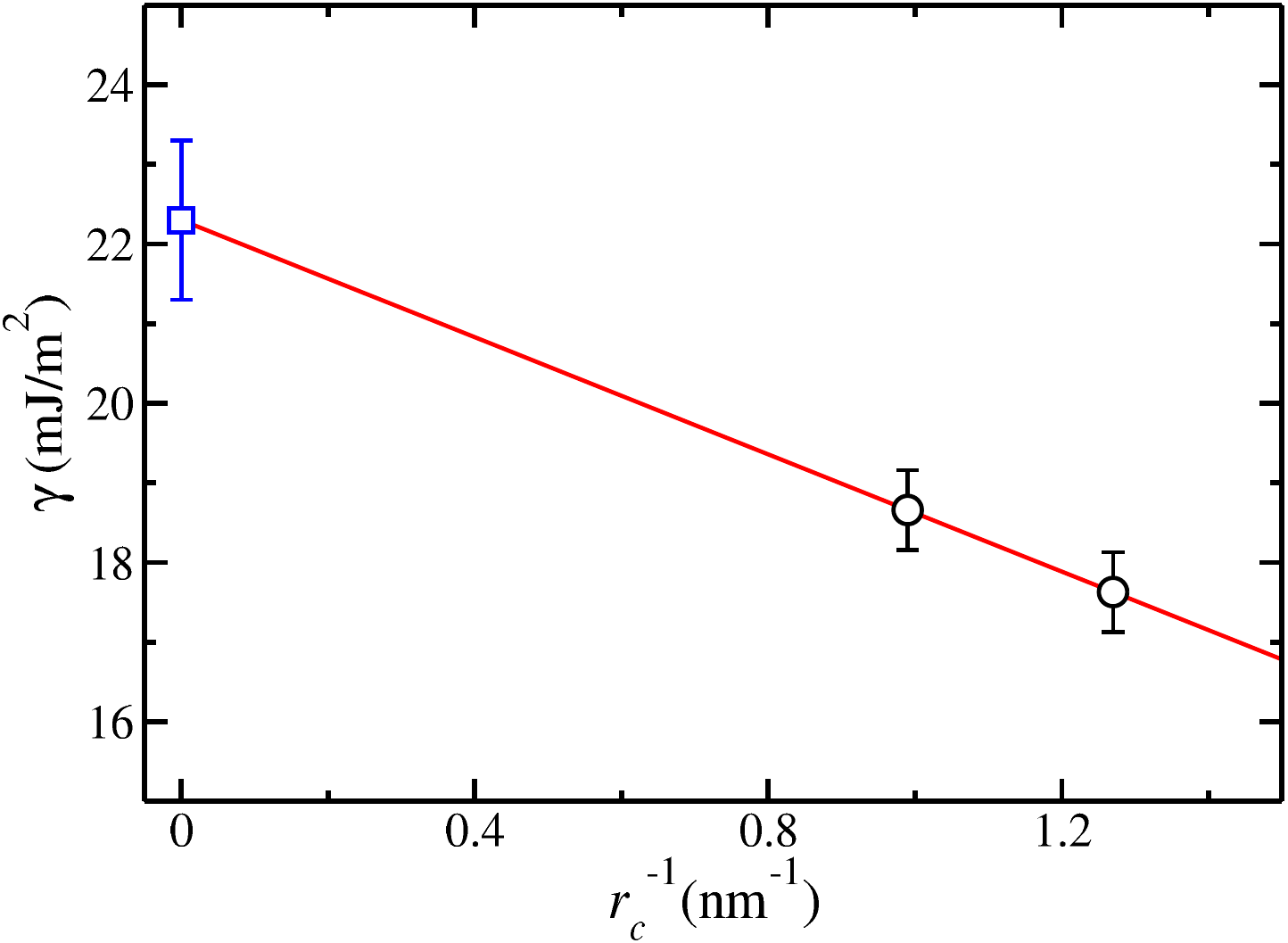}
\caption{CO$_{2}$ hydrate - solution interfacial free energy, $\gamma$, as a function of the inverse critical radius $r_{c}$  at $255\,\text{K}$ and $400\,\text{bar}$. Black circles represent the values found with Seeding for supersaturations $S=1$ (left) and $S=1.207$ (right). These values are obtained using the $\overline{q}_{3}-\overline{q}_{12}$ linear combination shown in Fig.~\ref{figure3} to get the number of molecules in the cluster. The red line corresponds to a linear fit of the $\gamma$ values obtained from Seeding. The blue square represents the extrapolated value of $\gamma$ ($22.3\,\text{mJ/m}^{2}$) obtained as $r_{c}\rightarrow\infty$ (planar interface).}
\label{gamma}
\end{figure}

\begin{table}
    \centering
    \caption{Nucleation rate of CO$_{2}$ hydrate in water, $J$, at $255\,\text{K}$, $400\,\text{bar}$, and supersaturations $S=1$ and $1.207$ using the Seeding methodology.}
    \begin{tabular}{lcccc}
    \hline
    \hline
        $S$  & \,\, & $1.0$ & \,\, &  $1.207$ \\
    \hline
        $N_{c}^{\text{H}_{2}\text{O}}$ & \,\, &  115 & \,\, & 55 \\
        $N_{c}^{\text{CO}_{2}}$ & \,\, & 20 & \,\, & 9.6  \\
        $Z$ & \,\,  & 0.077 & \,\, & 0.123  \\
        $\Delta G_{c}\,(k_{B}T)$& \,\, & 22.6  & \,\, & 13.06 \\
        $f^{+}_{\text{CO}_{2}}\,(\text{s}^{-1})$ & \,\, & 6.54 $\times 10^{8}$  & \,\, & 1.68 $\times 10^{9}$  \\
        $J\,(\text{m}^{-3}\,\text{s}^{-1})$ & \,\, & 2 $\times 10^{25}$ & \,\, & 1.36 $\times 10^{30}$  \\
        $\gamma\,(\text{mJ/m}^{2})$ & \,\, & 18.66  & \,\, & 17.63  \\
    \hline
    \hline        
    \end{tabular}
    \label{table_seeding}
\end{table}

\subsection{Interfacial free energy between the hydrate and the aqueous solution}

It is interesting to analyze in detail the expression leading to $J$ when using CNT (which is the expression used in the Seeding technique), and particularly 
Eqs.~\eqref{nucleation_rate_cnt_1} and \eqref{nucleation_rate_cnt_2} in the context of the CO$_{2}$ and CH$_{4}$ hydrates. It is important to recall again that the comparison between $J$ values for both hydrates is performed at the same pressure ($400\,\text{bar}$) and supercooling ($\Delta T=35\,\text{K}$). According to Eq.~\eqref{nucleation_rate_cnt_1}, $J$ is given by the product of a kinetic prefactor, $J_0$, and a free energy barrier within an exponential term. 
The comparison of $J_0$ for CH$_{4}$ and CO$_{2}$ hydrates shows that they are quite similar. They only differ in one order of magnitude but we must explain 30 orders of magnitude of difference. The Zeldovich factor of the CO$_{2}$ hydrate is twice that of methane, but the attachment rate is one-half so that the product of $Z$ and $f^{+}$ are almost identical in both cases. The density of the gas in the liquid phase is about one order of magnitude larger for CO$_{2}$ than for CH$_{4}$ (due to its higher solubility in water). Thus, the higher solubility of CO$_{2}$ in water affects the prefactor $_{0}$ (in the expression of $J$) by only one order of magnitude. Therefore, differences must come from the exponential free energy barrier, which has two components, $\Delta \mu_{\text{N}}$ and $N_c$. For $S=1$, 
$\Delta \mu_{\text{N}}$ amounts to $-2.26$ and $-2.42\,k_{B}T$  for the CO$_{2}$ and CH$_{4}$ hydrates, respectively. This goes in the right direction, as for a certain fixed value of $N_c$ the free energy barrier will be smaller for CO$_{2}$ than for CH$_{4}$. However, the difference does not seem so large to explain the difference in the nucleation rate. 
The difference in the nucleation rate comes from $N_c$ which contains $83$ molecules of methane but only $20$ of CO$_{2}$ at the same conditions. This is the key to understanding the differences: the critical cluster of the CO$_{2}$ hydrate is much smaller than that of the CH$_{4}$ hydrate. To analyze the physical origin of the difference let us consider
Eq.~\eqref{critical_nucleus_cnt} which describes the critical cluster size. Values of $\rho_s$ and $\Delta  \mu_{\text{N}}$ are quite similar for both hydrates. Therefore the key for the different behaviors must be on the value of the 
interfacial free energy $\gamma$ that moreover appears elevated to the third power.  

According to 
Montero de Hijes \emph{et al.},~\cite{Montero2019a} $\gamma$ should vary linearly with $1/r_{c}$. Particularly, they have found this relationship for several systems including the hard-sphere and Lennard-Jones simplified models, and more sophisticated force fields for water as the mW and TIP4P/Ice. This allows us to estimate the interfacial free energy of the corresponding planar solid-fluid interface from the knowledge of two values of $\gamma$ associated with two different critical CO$_{2}$ hydrate clusters. For further details, we refer to the reader to Fig.~2 of the work of Montero de Hijes \emph{et al.}~\cite{Montero2019a} Using Eq.~\eqref{critical_nucleus_cnt} one can calculate the values of
$\gamma$ as a function of the critical cluster radius for systems with $S=1$ and $S=1.207$ at $T=255\,\text{K}$ and $400\,\text{bar}$ and with these values extrapolate $\gamma$ to the planar interface ($r_{c}\to\infty$) as shown in  Fig.~\ref{gamma}. For $S=1$ the value of $\gamma$ for the CO$_{2}$ system is around $19\,\text{mJ/m}^{2}$ and the extrapolation to the hydrate-water planar interface yields $22.3\,\text{mJ/m}^{2}$. Notice that this value of the planar interface is not at the three-phase coexistence point but at the two-phase (hydrate-liquid) equilibrium at $250\,\text{K}$ and $400\,\text{bar}$ for the planar interface. See Fig.~13 of our last work on nucleation.\cite{Grabowska2022b}
However, for the CH$_{4}$ hydrate the value of $\gamma$ is of about $32\,\text{mJ/m}^{2}$ when $S=1$ and of about $39\,\text{mJ/m}^{2}$ for the planar hydrate-water interface.  Thus the higher nucleation rate of $J$ for the CO$_{2}$ hydrate as compared to the CH$_{4}$ hydrate arises from a lower value of $\gamma$ that decreases significantly the free energy barrier. Although it is almost impossible to present a molecular explanation one could argue that when the composition of the fluid phase is more similar to that of the hydrate (which has a molar fraction of the gas molecule of $8/54 = 0.148$)  the interfacial free energy becomes smaller. The higher values of $\gamma$ for the CH$_{4}$ hydrate-water interface would arise from a larger difference in composition between the aqueous phase and the hydrate. Thus the higher solubility of CO$_{2}$ in water would affect the nucleation rate not in the kinetic prefactor, which only contributes to the different in one order of magnitude, nor in the value of $\Delta \mu_{\text{N}}$ but on decreasing significantly the value of $\gamma$. 

There is an additional interesting observation. The value of $\gamma$ of the hydrate-water planar interface for the CO$_{2}$ systems seems to increase with temperature along the two-phase coexistence line.  In fact, for $255\,\text{K}$ the estimated value is $22\,\text{mJ/m}^2$ whereas 
$\gamma$ is around $30(2)\,\text{mJ/m}^{2}$ at $T_3=290\,\text{K}$ at this pressure according to previous calculations by some of us using the mold integration host and guest methodology.\cite{Algaba2022b, Zeron2022a}

\begin{figure}
\centering
\includegraphics[width=0.9\columnwidth]{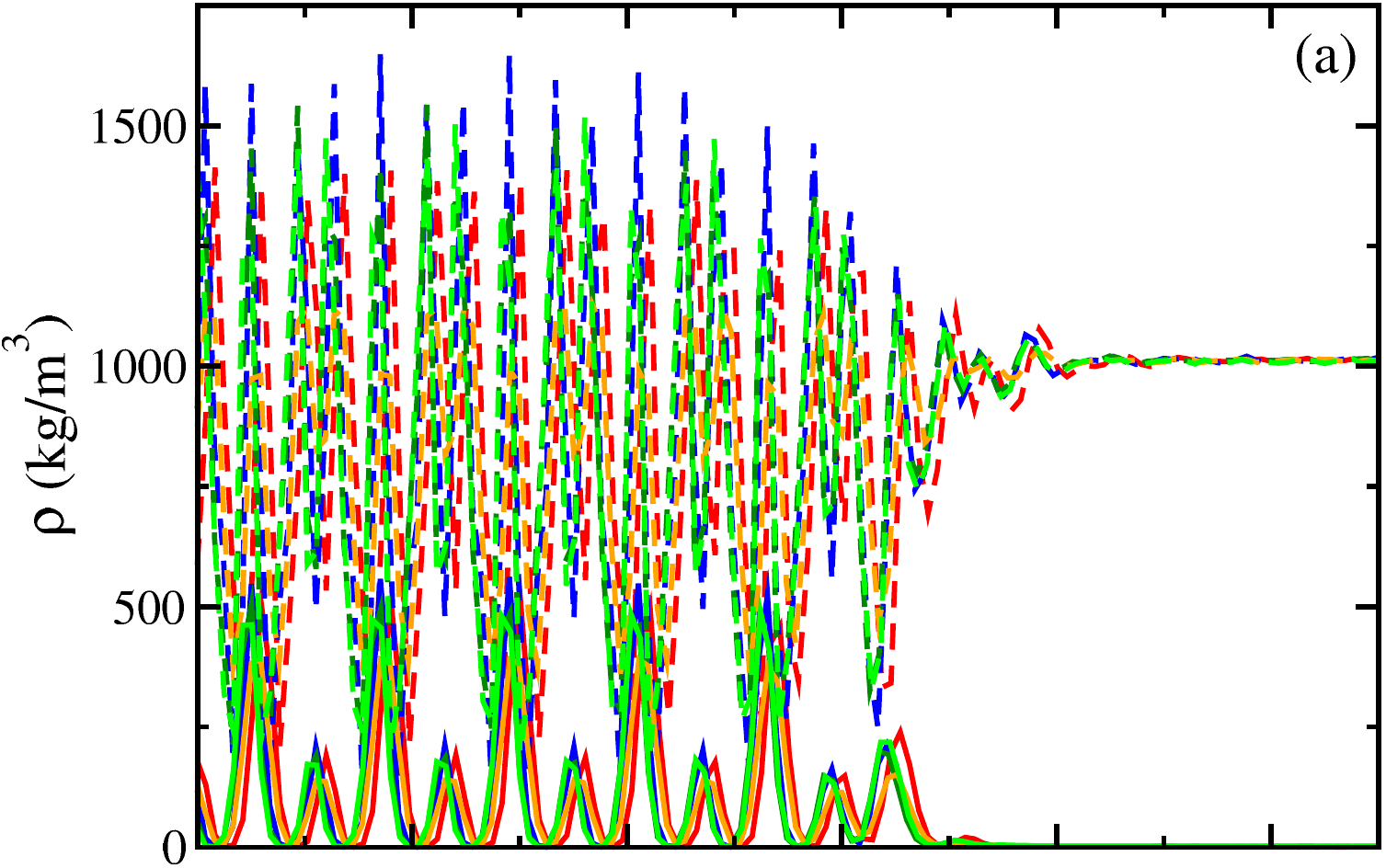}\\
\includegraphics[width=0.9\columnwidth]{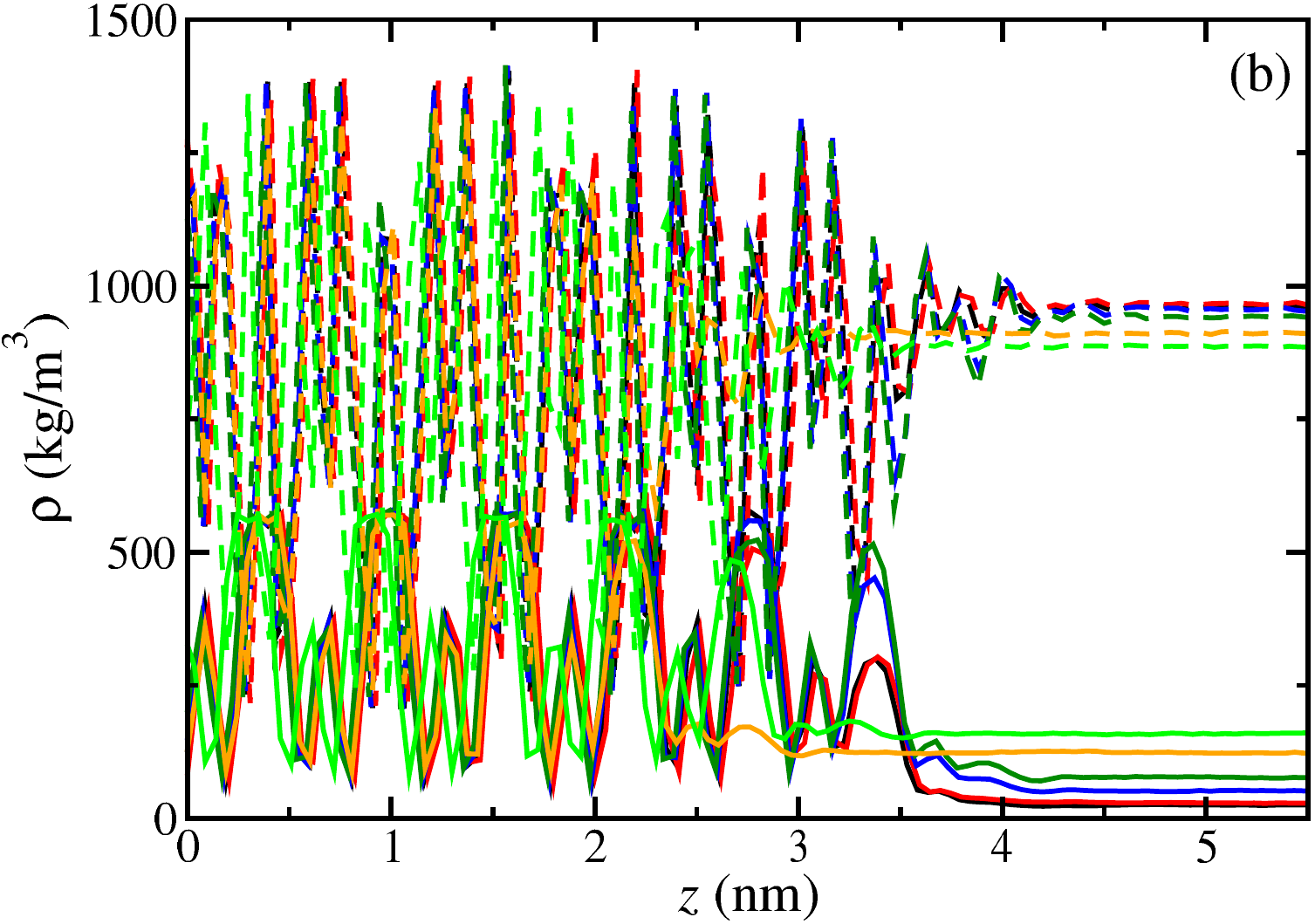}
\caption{Simulated equilibrium density profiles of methane and carbon dioxide (continuous curves in both panels) and water (dashed curves), $\rho(z)$, across the hydrate–liquid interface as obtained from MD anisotropic $NPT$ simulations at $400\,\text{bar}$ and $250$ (black), $260$ (red), $270$ (blue), $280$ (dark green), $290$ (orange), and $295\,\text{K}$ (light green). Panel (a) corresponds to the CH$_{4}$ hydrate-liquid interface and panel (b) to the CO$_{2}$ hydrate-liquid interface.}
\label{fig:HL}
\end{figure}

It is also useful to inspect the density profiles of CO$_{2}$ and water at the CO$_{2}$ hydrate - water interface and to compare with those corresponding to the CH$_{4}$ hydrate -water interface. Fig.~\ref{fig:HL} shows the density profiles of water and CH$_{4}$ molecules in panel (a) and of water and CO$_{2}$ in panel (b). Results were obtained in our previous works.~\cite{Grabowska2022a,Algaba2023a} Notice that results for the CO$_{2}$ hydrate were already presented in Fig.~6 of the work of Algaba \emph{et al.}~\cite{Algaba2023a} In both cases, results were obtained from anisotropic $NPT$ simulations at $400\,\text{bar}$ and temperatures ranging from $250$ to $295\,\text{K}$. As can be observed, the profiles of CO$_{2}$ and water in the hydrate phase and near the interface, shown in panel (b), exhibit the usual behavior expected in solid-fluid coexistence. Notice that the density profiles of CH$_{4}$ and water near the corresponding interface, presented in panel (a), show the same structural order due to the presence of the hydrate phase. There are some differences in behavior between the excess concentration of CO$_{2}$ on the surface compared to CH$_{4}$. First, we can see that the outwards most peaks of the two hydrate phases (at around $3-3.5\,\text{nm}$) are rather different, with the peak for CO$_{2}$ being broader. Secondly, there is a ``tail'' for the CO$_{2}$ profiles, which decay significantly slower than the CH$_{4}$ case. The tail of the CO$_{2}$ distributions stabilizes only at between $4-4.5\,\text{nm}$. The results of Fig.~\ref{fig:HL} seem to suggest an excess of CO$_{2}$ at the water-hydrate interface (although the rigorous determination of the adsorption of the gas at the hydrate-water interface is left to future work). This may provide a mechanism that further decreases the free energy between the hydrate and the CO$_{2}$ aqueous solution.

Finally, it would be interesting in this context to estimate the empty hydrate - water interfacial free energy to compare with the values obtained here and in previous works~\cite{Algaba2023a,Algaba2022b,Zeron2022a,Romero-Guzman2023a} for the CH$_{4}$ and CO$_{2}$ hydrates. However, empty structures of hydrates, including sI, sII, and sH, are usually called virtual ices. According to Conde \emph{et al.},~\cite{Conde2009a} the empty hydrates sII and sH appear to be the stable solid phases of water at negative pressures. Consequently, the sI and other virtual ices do not enter the phase diagram shown in Fig.~5 of the work of Conde \emph{et al.}~\cite{Conde2009a} In other words, no pressure or temperature conditions exist at which these structures have lower chemical potential than I$_{\text{h}}$, sII, or sH crystallographic structures. Thus there is a high risk for the growth of another phase of ice from a template of sI (using for instance the Mold Integration technique~\cite{Espinosa2014a,Espinosa2016a,Algaba2022b}) and that would prevent the determination of the value of $\gamma$ for the sI-water interface. This issue should be examined in greater detail in future work.

\subsection{Nucleation along the $S = 1$ curve}

The value of $J$ at 255 K for $S=1$ is of the order of $10^{25}/(\text{m}^3 \text{s})$. Nucleation can be observed spontaneously in BF runs when the nucleation rate is larger than $10^{29}/(\text{m}^3 \text{s})$ with current computational resources. 
Therefore, it seems likely that nucleation can be observed in BF runs at $S = 1$ if we move to lower temperatures (thus increasing the driving force). This is of particular interest as nucleation studies in experiments are usually performed along the $S = 1$ curve (with the solution in contact with a gas reservoir \cite{Kashchiev2002b}).

We performed BF runs at 245 and $250\,\text{K}$ (and $400\,\text{bar}$) at the corresponding  CO$_{2}$  saturation  concentration. 
These states are represented as red diamonds in Fig.~\ref{fig:pd} (note that they are located on the $S=1$ red line). Details on these simulations are given in Table \ref{table_bf}. As we have already mentioned, we used two types of simulation setups for this study: a homogeneous CO$_2$ saturated bulk solution (denoted as ``one-phase system'' in Table \ref{table_bf}) and a saturated solution in contact with a fluid CO$_2$ reservoir (denoted as ``two-phase system'' in Table \ref{table_bf}).  
We focus first on the one-phase system and analyze later on the comparison between both setups. 
We used isotropic $NPT$ runs for the one-phase systems. 
As not all trajectories were successful in nucleating the solid phase, we used the method of Walsh et al. \cite{Walsh2011a}, described previously in the manuscript when discussing the BF runs for $S = 1.207$, to determine the nucleation rate.  We obtain a nucleation rate of the order of $10^{31}\, /\text{(m}^{3}\text{s)}$ for $245\,\text{K}$ and of $10^{29}\, /\text{(m}^{3}\text{s)}$ for $250\,\text{K}$. These nucleation rates are fully consistent with the $J=10^{25}\,\text{m}^{-3}\text{s}^{-1}$ obtained in this work for $S = 1$ at $255\,\text{K}$ via Seeding (see Section \ref{sec:seeding}). The $J$ values obtained from BF simulations along the $S = 1$ line (red circles) are compared with that
obtained via Seeding for $S = 1$ (green triangle) in Fig.~\ref{fig:s1}a.

\begin{table}
    \centering
    \caption{Simulation details and results obtained using BF simulations at $250$ and $245\,\text{K}$, both at $400\,\text{bar}$ and supersaturation $S = 1$ in one- and two-phase systems.}
    \begin{tabular}{lcccc}
    \hline
    \hline
         & \,\, & one-phase system & \,\, &  two-phase system \\
    \hline
        $T\,(\text{K})$ & \,\,  &  \multicolumn{3}{c}{$250$} \\
        \hline
        $N^{\text{H}_{2}\text{O}}$         & \,\,  &  6524 & \,\, & 7200   \\
        $N^{\text{CO}_{2}}$         & \,\,  & 606   & \,\, & 3444    \\
        $x_{\text{CO}_2}$           & \,\,  & 0.085 & \,\, & ~0.085  \\
        box dim. (nm$^3$)   & \,\,  & $6.06\times6.06\times6.06$  & \,\, & $9.07\times6.25\times7.41$ \\
        liquid dim. (nm$^3$)  & \,\,  & $6.06\times6.06\times6.06$ & \,\, & $5.77\times6.25\times7.41$  \\
        $V_{liq}$ (nm$^3$)  & \,\,  & 222    & \,\, & 267  \\
        $n_{runs}$           & \,\,  & 12     & \,\, & 12  \\
        $n_{nucl}$           & \,\,  & 4      & \,\, & 4     \\
        $t_{nucl}\,(\text{ns})$      & \,\,  & 730    & \,\, & 245   \\
                             & \,\,  & 1310   & \,\, & 285   \\
                             & \,\,  & 1570   & \,\, & 480   \\
                             & \,\,  & 1700   & \,\, & 1510  \\
        $t_{total}\,(\text{ns})$    & \,\,  & 21310  & \,\, & 18520 \\
        $\rho_{L}^{\text{CO}_{2}}\,(\text{m}^{-3})$ & \,\,  & $2.7\times 10^{27}$ & \,\, & $2.7\times 10^{27}$  \\
         $J\,(\text{m}^{-3}\,\text{s}^{-1})$     & \,\,  & $8.45\times 10^{29}$  & \,\, & $8.09\times 10^{29}$ \\
    \hline
        $T\,(\text{K})$  & \,\,  &  \multicolumn{3}{c}{$245$}\\
        \hline
            $N^{\text{H}_{2}\text{O}}$         & \,\,  &  2400 & \,\, & 2400   \\
        $N^{\text{CO}_{2}}$         & \,\,  & 240   & \,\, & 1148    \\
        $x_{\text{CO}_2}$           & \,\,  & 0.09 & \,\, & ~0.09  \\
        box dim. (nm$^3$)    & \,\,  & $4.35\times4.35\times4.35$  & \,\, & $9.19\times6.25\times2.47$ \\
        $V_{liq}$ (nm$^3$) & \,\,  & 82.5    & \,\, & 82.5  \\
        $n_{runs}$           & \,\,  & 2     & \,\, & 2  \\
        $n_{nucl}$           & \,\,  & 2      & \,\, & 2     \\
        $t_{nucl}\,(\text{ns})$      & \,\,  & 640    & \,\, & 300   \\
                             & \,\,  & 550   & \,\, & 230   \\
        $t_{total}\,(\text{ns})$    & \,\,  & 1190  & \,\, & 530 \\
                $\rho_{L}^{\text{CO}_{2}}\,(\text{m}^{-3})$ & \,\,  & 2.9 $\times 10^{27}$ & \,\, & 2.9 $\times 10^{27}$ \\
                        $J\,(\text{m}^{-3}\,\text{s}^{-1})$     & \,\,  & 2.02 $\times 10^{31}$  & \,\, & 5.78 $\times 10^{31}$ \\
    \hline      
    \hline
    \end{tabular}
    \label{table_bf}
\end{table}

In the following sections, we describe how we combine our nucleation studies at $S=1$ and three different 
temperatures (245, 250, and 255 K) with a recent calculation of $\gamma$ between the solid and the solution
at the three-phase coexistence temperature \cite{Algaba2022b} to get an estimate of the whole $J(T)$ curve along the $S=1$ line. 

\subsubsection{$\gamma$ along $S = 1$}
\begin{figure}
\centering
\includegraphics[width=0.9\columnwidth]{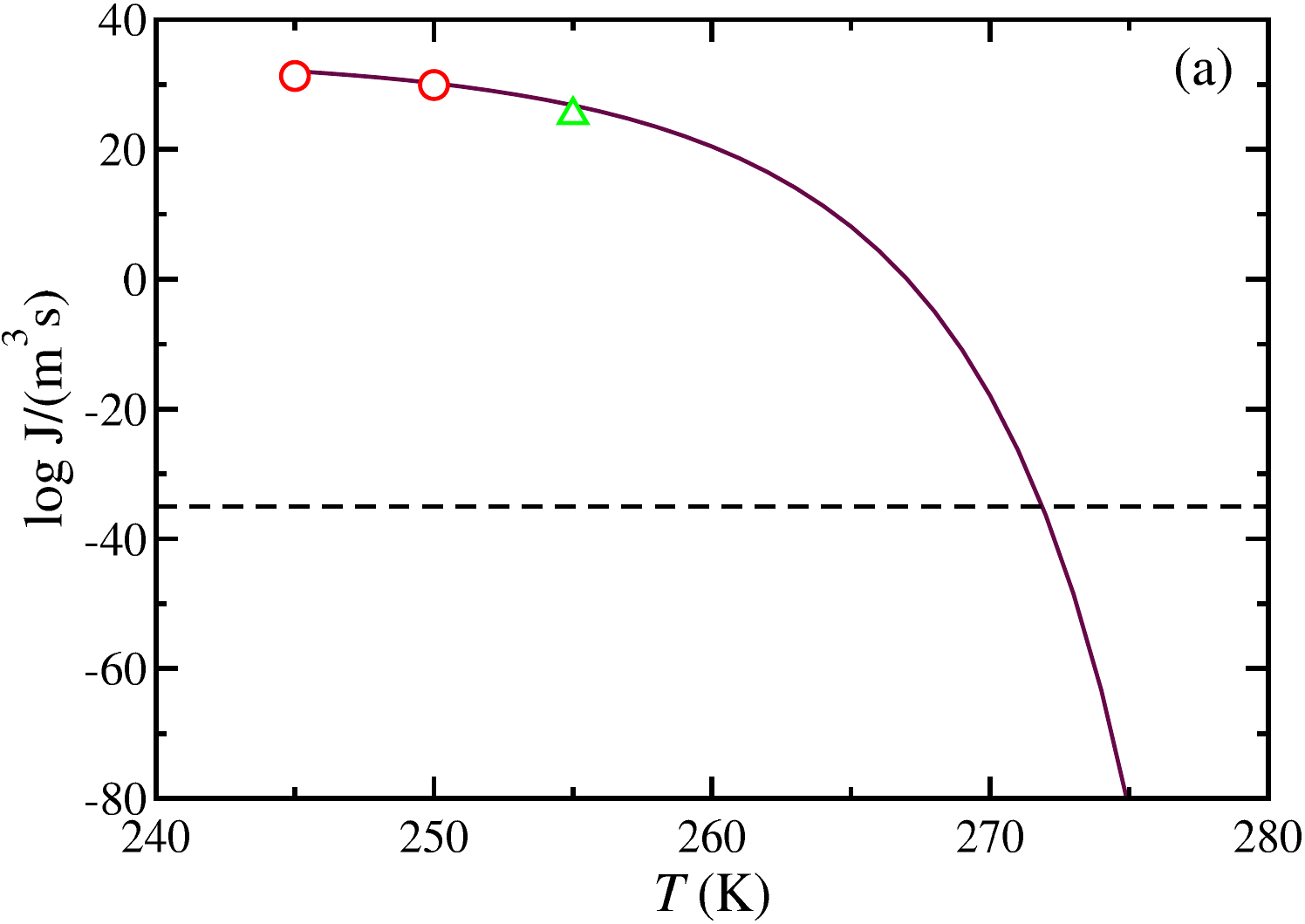}\\
\includegraphics[width=0.865\columnwidth]{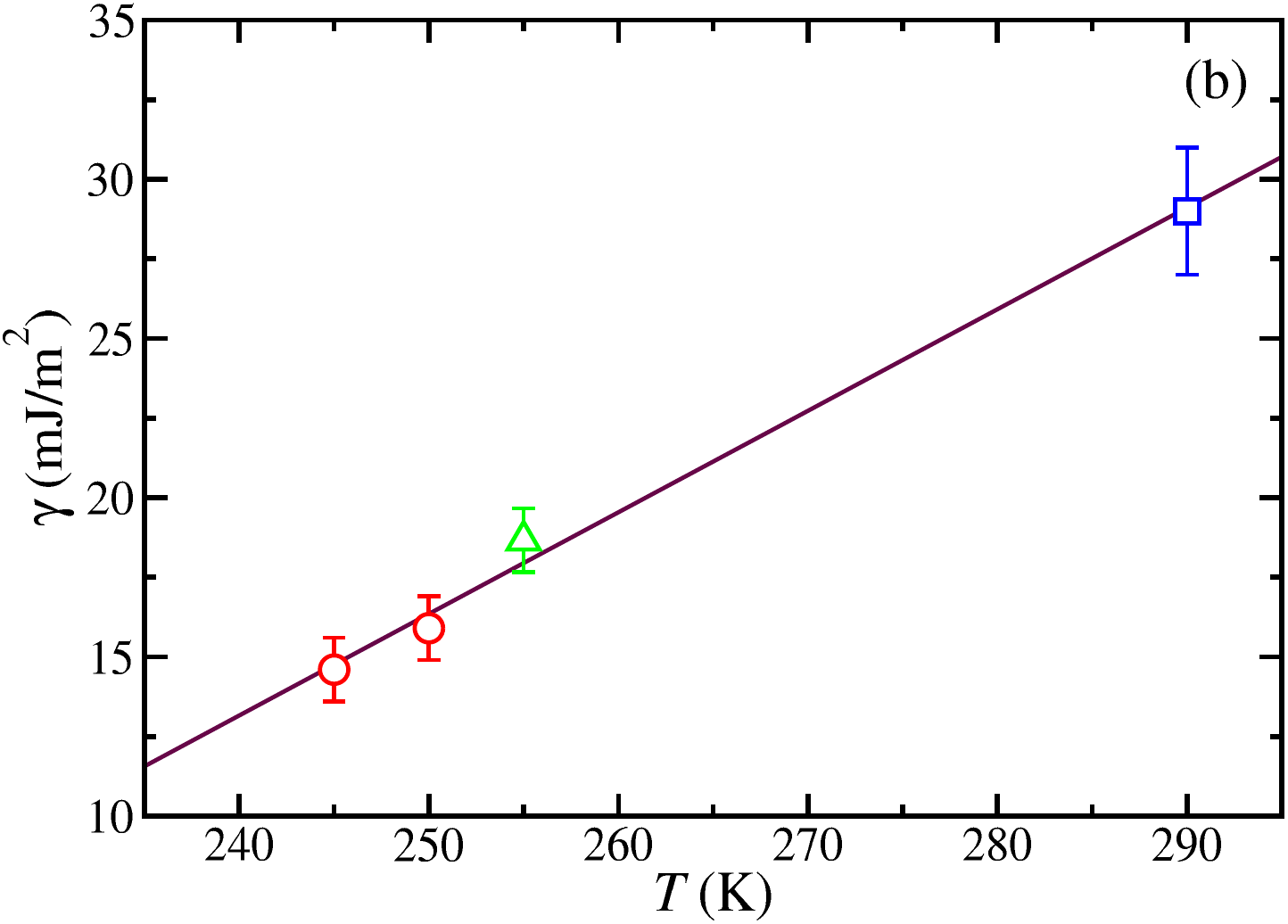}
\caption{CO$_{2}$ hydrate nucleation rate, $J$, (a) and CO$_{2}$ hydrate-water interfacial free energy, $\gamma$, (b), as functions of temperature along the $S=1$ curve. Red circles and green triangles are BF and Seeding results obtained in this work, respectively. The blue square is the CO$_{2}$ hydrate-water interfacial free energy at coexistence conditions obtained by Algaba \emph{et al.}~\cite{Algaba2022b} through the Mold Integration.~\cite{Espinosa2014a} Continuous curve in (a) is obtained using simulation data via the CNT approach and 
line in (b) is a linear fit of simulation data. The curve of $J$ as a function of time is obtained using the $\gamma(T)$ dependence found in (b). The dashed horizontal line in (a) corresponds 
to an ``unachievable'' nucleation rate given by one nucleus per universe age and hydrosphere volume.}
\label{fig:s1}
\end{figure}

To estimate $J$ along $S = 1$ we need first to know how $\gamma$ varies along $S = 1$, given that the nucleation barrier can be obtained from $\gamma$ through Eq.~\eqref{eq:barrier}.
We already have a value of $\gamma$ from Seeding at $S=1$ and $255\,\text{K}$ ($18.7\,\text{mJ/m}^{2}$, depicted with a green triangle in Fig.~\ref{fig:s1}b).  Recently, some of us estimated $\gamma$ at the $T_3$ (the temperature where solid, solution and
CO$_2$ reservoir coexists,\cite{Algaba2022b} indicated by an maroon circle in Fig.~\ref{fig:pd} for $S=1$). The value found was $29(2)\,\text{mJ/m}^{2}$ at $287\,\text{K}$. The $T_3$ value was later refined to $290\,\text{K}$.\cite{Algaba2023a} We assume here that the value of $\gamma$ found in the work of Algaba and collaborators~\cite{Algaba2022b} at $287\,\text{K}$ is valid for the updated  $T_3$ of $290\,\text{K}$ (the temperature difference between both $T_3$ estimates is small). The value of $\gamma$ at $T_3$ is shown as a blue square in Fig.~\ref{fig:s1}b). 

\begin{figure}
\centering
\hspace*{-0.2cm}
\includegraphics[width=0.9\columnwidth]{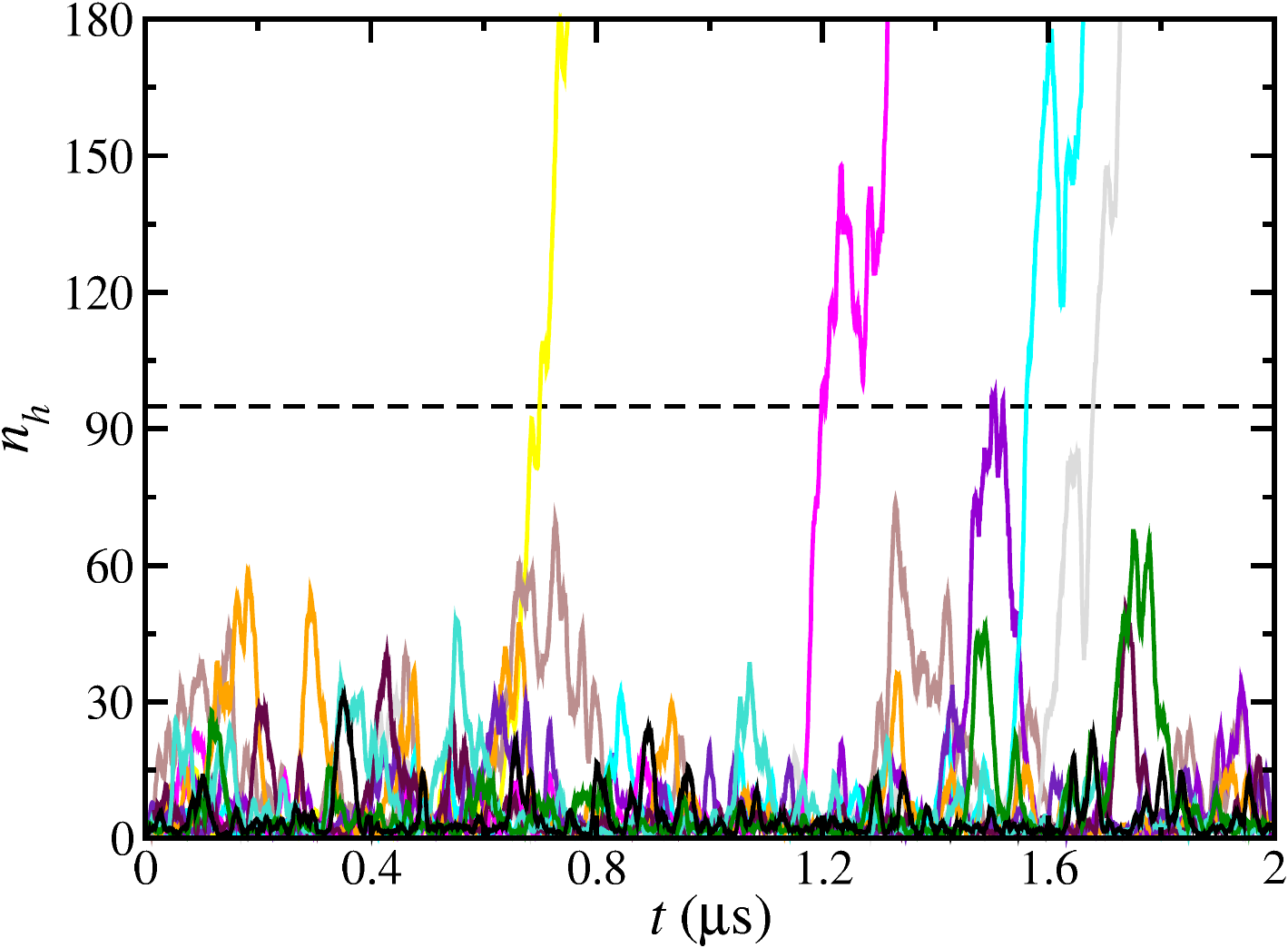}
\caption{Number of water molecules in the largest
cluster of the CO$_{2}$ hydrate, $n_{h}$, as a function of time, for different bulk simulations at $S=1$ and $250\,\text{K}$. CO$_{2}$ hydrate growth is observed in four trajectories. The dashed horizontal line, which highlights the largest sub-critical cluster that emerged in all simulations, is our estimate for $N_c^{\text{H}_2\text{O}}$ using the $\overline{q}_{3}-\overline{q}_{12}$ linear combination of the local bond order parameters (95 in the present example).}
\label{fig:spont250s1}
\end{figure}

We now try to get an estimate of $\gamma$ from the two BF simulation studies performed at 245 and $250\,\text{K}$. To do that we use Eq.~\eqref{nucleation_rate_cnt_1} to get 
$\Delta G_c$ from $J$ and, then, Eq.~\eqref{eq:barrier} to obtain $\gamma$ from $\Delta G_c$ (the CO$_2$ density in the solid phase used for these calculations is $\rho_{S}^{\text{CO}_{2}}=4.7\times 10^{27}\,\text{m}^{-3}$ for both temperatures). 
The first of these two steps requires an estimate of $J_0$.
To estimate $J_0$ we need $f^+$. We use the fact that $f^+$ is proportional to the CO$_2$ diffusion coefficient, $D_{\text{CO}_2}$, and to $N_c^{2/3}$,\cite{Espinosa2016c} 
to obtain $f^+$ at $245$ and $250\,\text{K}$ from the $f^+$ calculated at 
$255\,\text{K}$.  This requires computing 
$D_{\text{CO}_2}$ at $245$, $250$, and $255\,\text{K}$ and estimating $N_c$ in the BF runs. 
The CO$_2$ diffusion coefficients we get from $NPT$ simulations of 
the aqueous solutions are 1.6, 2.2, and 3.0 $\times$ $10^{-11}\,\text{m}^{-2}\text{s}^{-1}$ for 245, 250, and $255\,\text{K}$
respectively. 
To estimate $N_c$  we identify the largest cluster 
that appears during the induction period previous to 
hydrate growth (see Fig.~\ref{fig:spont250s1} ).
In this way, we get 42 and 95 water molecules in the 
critical cluster at 245 and $250\,\text{K}$, respectively, which are values fully consistent 
with $N_c^{\text{H}_2\text{O}}$= 115 obtained with Seeding at $255\,\text{K}$.  
While this estimate of $N_c$ might not be accurate, the final value of $\gamma$ is not significantly influenced by this inaccuracy, as we argue further on.
The $|\Delta \mu_{\text{N}}|$ factor in $J_0$ is taken from our previous work using the route 4.~\cite{Algaba2023a}  
We get $|\Delta \mu_{\text{N}}|=$  2.98 and $2.59\,k_{B}T$ at 245 and $250\,\text{K}$, respectively. 
With these ingredients we obtain the $\gamma$ estimates shown 
in Fig.~\ref{fig:s1}b as red circles ($14.6$ and $16.0\,\text{mJ/m}^{2}$ at $245$ and $250\,\text{K}$ respectively). 

The $\gamma$ estimates from BF simulations (red circles), 
from Seeding simulations (green triangles), and from Mold Integration (blue square) are fully consistent 
among each other and can be fitted quite nicely to a straight line (maroon line). $\gamma$ increases with temperature along 
the $S=1$ line roughly at a rate of $1\,\text{mJ/m}^{2}$ every $3\,\text{K}$. 

Interestingly, BF simulations yield a $\gamma$ value less sensitive to the order parameter than the Seeding method. In Seeding, Eq.~\eqref{critical_nucleus_cnt} is used to infer $\gamma$ from the $N_c$ value obtained in seeded simulations, which has a strong dependence on the chosen order parameter. In contrast, in the BF approach, $N_c$ is used for estimating the kinetic pre-factor. As the natural logarithm of this pre-factor is taken to calculate $\Delta G_c$ (from which $\gamma$ is then obtained via Eq.~\eqref{eq:barrier}), the influence of $N_c$ on the final $\gamma$ value is relatively minor.
To illustrate this, let us consider the impact of doubling the cluster size in the calculation 
of $\gamma$. In Seeding ($255\,\text{K}$), $\gamma$ would significantly increase from 18.7 to $23.5\,\text{mJ/m}^{2}$. However, in BF simulations, the changes are much smaller: from 14.6 to $14.9\,\text{mJ/m}^{2}$ at 245 K, and from 16.0 to $16.2\,\text{mJ/m}^{2}$ at 250 K. In conclusion: (i) BF simulations provide an estimate of $\gamma$ less influenced by the choice of order parameter than Seeding; (ii) the way in which we obtain $N_c$  from BF simulations is good enough to get a reliable estimate of $\gamma$. 

\subsubsection{$J$ along $S=1$}

 Using the linear fit of $\gamma(T)$ shown in Fig.~\ref{fig:s1}b, we can obtain 
$\Delta G_c$ at any temperature using Eq.~\eqref{nucleation_rate_cnt_1} with the 
$|\Delta \mu_{\text{N}}|$ obtained in our previous work (route 4).~\cite{Algaba2023a} We use
the following fit for the chemical potential difference: $|\Delta \mu_{\text{N}}|/(k_BT) = -3.02 \times 10^{-4} \, T^2 + 0.228 \, T - 40.7$. With $\Delta G_c$ and Eq.~\eqref{nucleation_rate_cnt_2} we can estimate $J$ at any temperature provided that we have $J_0$ as well. 
This requires having $f^{+}$ at any $T$ (see Eq.~\eqref{nucleation_rate_cnt_1}). 
For that purpose, 
we use again the fact that $f^{+}\propto N_c^{2/3} D_{\text{CO}_2}$. \cite{Espinosa2016c} On the one hand, through $NPT$ simulations of the saturated aqueous solution 
at different temperatures, we got the 
following fit to obtain $D_{\text{CO}_2}$ at any temperature: $\ln  [D/(\text{m}^2/\text{s})]= -0.0011 \, T^2 + 0.6846 \, T - 124.99$.
On the other hand, $N_c$ can be obtained at any $T$ using $\Delta G_c=N_{c}|\Delta \mu|/2$ according to Eq.~\eqref{driving_force_cnt}. 
The missing factors to complete the calculation of $J_0$ (and of $J$) are the Zeldovich factor $Z$, 
that can be easily computed through $N_C$ and $|\Delta \mu_{\text{N}}|$ (Eq.~\eqref{zeldovich}), and $\rho_{L}^{\text{CO}_{2}}$ which is trivially obtained in $NPT$ simulations. 
With these ingredients, we can draw the maroon curve in Fig.~\ref{fig:s1}a that predicts the trend of the nucleation rate at $S=1$. 

Unfortunately, to the best of our knowledge, there is no experimental data to compare these simulation predictions with. In homogeneous ice nucleation, rates of the order of $10^{2} - 10^{16}\,\text{m}^{-3}\,\text{s}^{-1}$ (with microdroplets) and higher (with nanodroplets) are experimentally accessible.~\cite{espinosa2018homogeneous} Our predictions indicate that such rates occur at temperatures below 266 K (beyond 25 K supercooling). Therefore, we hope that simulations and experiments of homogeneous hydrate nucleation can be compared in the future, 
as they were for the case of ice nucleation. \cite{espinosa2018homogeneous}

\subsubsection{Bulk versus surface nucleation}

The dashed horizontal line in Fig.~\ref{fig:s1}a indicates the order of magnitude of an unachievable 
nucleation rate: that corresponding to 1 nucleus formed in the volume of the hydrosphere 
and in the age of the universe. 
Our CNT fit (maroon curve) predicts that this unattainable rate occurs at about $272\,\text{K}$ (around $20\,\text{K}$ below $T_3$). 
Therefore, any crystallization event at a supercooling lower than $20\,\text{K}$ must
be heterogeneous (the difficulty of observing homogeneous
nucleation was also highlighted in a simulation study of
methane hydrates).~\cite{Knott2012a} In most experiments, hydrate crystallization typically occurs at supercooling conditions of less than $20\,\text{K}$.~\cite{barwood2022extracting,metaxas2019gas,Maeda2018}
Such low supercooling suggests that the nucleation of hydrates is not homogeneous in the real world. 
Although experiments do not provide molecular insight into the nucleation step, 
it is commonly believed that nucleation occurs at the gas-solution interface, perhaps
assisted by impurities, the glass-solution contact line,~\cite{maeda2015nucleation} or aided by an increased concentration of the hydrate formed near
the interface.~\cite{Warrier2016a}

To investigate the latter hypothesis we compare BF simulation runs 
at 245 and $250\,\text{K}$ performed in two-phase systems (where the solution is in contact with a CO$_2$ reservoir) 
with those run in one-phase systems that have been already presented (without a bulk aqueous solution having the equilibrium CO$_2$ saturation 
concentration). 
In the two-phase simulations, the details of which are reported in Table \ref{table_bf}, we used $NP_z\mathcal{A}T$ runs. 
Obviously, the volume used to calculate the nucleation rate is only that of the aqueous phase in two-phase systems. 
As reported in Table~\ref{table_bf}, both simulation setups give 
the same nucleation rate for both temperatures (within less than half an order of magnitude). 
Therefore, the CO$_2$-solution interface does not have any effect on hydrate nucleation, 
at least at 245 and $250\,\text{K}$. 
However, there could be a crossover between homogeneous
and heterogeneous nucleation as the temperature increases (as is the case for crystallization of hard spheres with density)~\cite{espinosa2019heterogeneous} that could explain nucleation events at low supercooling.
More research is needed to identify the nucleation path in mild supercooling conditions,
where hydrate formation is experimentally observed. 

\section{Conclusions}

In this work, we have calculated the homogeneous nucleation rate of CO$_{2}$ hydrate at $400\,\text{bar}$ and $255\,\text{K}$ ($35\,\text{K}$ of supercooling) using Classical Nucleation Theory and Seeding simulations. For supersaturated systems (i.e., $S=1.207$ and $S=1.268$ ) the nucleation rate can be obtained from BF simulations. 
Since the results of Seeding depend on the choice of the order parameter we tested that a combination of  $\overline{q}_{3}$ and $\overline{q}_{12}$ is able to distinguish in an efficient way molecules of water belonging to the liquid or to the hydrate with a mislabeling of about 0.02$\%$. By using this combination of order parameters in Seeding runs with $S=1.207$ we confirmed that it provides an estimate of $J$ in full agreement with that obtained from BF runs. In other words, the selected order parameter allows a satisfactory estimate of the radius of the solid critical cluster at the surface of tension. 

After checking the adequacy of the order parameter we implemented the Seeding technique (in a system having two phases) for $S=1$ at $255\,\text{K}$ and $400\,\text{bar}$.
We obtained a size of 115 molecules of water for the critical cluster 
and a value of  of $10^{25}/(\text{m}^3 \text{s})$ for the nucleation rate. This is about 30 orders of magnitude larger than the value obtained in our previous work for the methane hydrate at the same pressure and supercooling. The higher solubility of CO$_{2}$ is not sufficient to explain such an enormous difference. We identify that the key is a much lower value of $\gamma$ for the CO$_{2}$ hydrate-water interface when compared to that of the CH$_{4}$ hydrate-water interface, and speculate that the value of
$\gamma$ in these systems could be lower when the composition of the solution becomes closer to the composition of the hydrate. The interfacial free energy of the CO$_{2}$ hydrate at $S=1$ was of about $19\,\text{mJ/m}^2$ as compared to the value of $29\,\text{mJ/m}^2$ obtained in our previous work for the methane hydrate. 
This means that at the same supercooling, the nucleation rate of CO$_{2}$ hydrate is 30 orders of magnitude higher than the estimation found in our last work of nucleation rate of methane hydrate\cite{Grabowska2022b} and $20$ orders of magnitude higher than the nucleation rate of ice I$_{\text{h}}$ which at this pressure and supercooling is of around $J_{\text{I}_{\text{h}}} = 10^{5}/(\text{m}^{3}\text{s})$.\cite{bianco2021anomalous}

We found that the energy to create the planar hydrate-liquid interface is $\gamma =22.3\,\text{mJ/m}^2$ at $255\,\text{K}$ and $400\,\text{bar}$, this suggests that the interfacial free energy for a planar interface should increase as the system moves along the two-phase curve from this supercooling temperature to the $T_{3}$, where $\gamma$ is around $30(2)\,\text{mJ/m}^2$ according to experiments\cite{Uchida1999a,Uchida2002a,Anderson2003a,Anderson2003b} and our previous calculations using the mold integration host and guest methodology.\cite{Algaba2022b, Zeron2022a}

 Finally, we have shown that BF runs in a two-phase system can indeed be performed to nucleate the hydrate at 245 and $250\,\text{K}$ to obtain $J$ when $S=1$ at these temperatures. Comparison of the value of $J$ from simulations using two phases with a system having just one phase reveals that the water-CO$_{2}$ interface does not enhance the nucleation rate so that at least for temperatures below $255\,\text{K}$ the nucleation is homogeneous and there is not an enhancement of the nucleation rate due to heterogeneous nucleation at the water-CO$_{2}$ interface. 
 However, there could be a crossover to heterogeneous nucleation at higher temperatures so that it is the main path to nucleation when closer to the equilibrium temperature $T_3$.
 Finally, we estimate the value of $J$ along the $S=1$ curve concluding that homogeneous nucleation could indeed be determined experimentally at this pressure for supercooling 
 larger than 25 $\text{K}$. Our simulations
 predict that homogeneous nucleation is not
 viable for supercooling lower than 20 K. Therefore,
 nucleation must be heterogeneous in typical experiments
 where hydrate formation is observed at low supercooling.

\section*{Acknowledgements}
This work was financed by Ministerio de Ciencia e Innovaci\'on (Grant No.~PID2021-125081NB-I00 and PID2024-158030NB-I00), Junta de Andalucía (P20-00363), and Universidad de Huelva (P.O. FEDER UHU-1255522, FEDER-UHU-202034, and EPIT1282023), all six cofinanced by EU FEDER funds. We greatly acknowledge the RES resources provided by the Barcelona Supercomputing Center in Mare Nostrum to FI-2023-2-0041. The authors also acknowledge Project No.~PID2019-105898GB-C21 of the Ministerio de Educaci\'on y Cultura. We also acknowledge access to supercomputer time from RES from project FI-2022-1-0019. J. G. gratefully acknowledges Polish high-performance computing infrastructure PLGrid (HPC Center: ACK Cyfronet AGH) for providing computer facilities and support within computational grant no. PLG/2024/017195. Part of the computations were carried out at the Centre of Informatics Tricity Academic Supercomputer \& Network. C.~Vega, E.~Sanz, and S.~Blazquez acknowledge the funding from project PID2022-136919NB-C31 of Ministerio de Ciencia e Innovacion.

\section*{Author declarations}

\section*{Conflicts of interest}
The authors have no conflicts to disclose.

\section*{Author contributions}

\textbf{Iván M. Zerón}: Conceptualization (equal); Data curation (lead); Formal analysis (lead); Investigation (equal); Writing – original draft (equal); Writing – review \& editing (equal). \textbf{Jesús Algaba}: Conceptualization (lead); Data curation (equal); Investigation (equal); Methodology (equal); Supervision (equal); Writing – review \& editing (equal). \textbf{José Manuel Míguez}: Conceptualization (equal); Data curation (equal); Investigation (equal); Methodology (equal); Writing – review \& editing (equal). \textbf{Joanna Grabowska}: Investigation (equal); Methodology (equal); Writing – review \& editing (equal). \textbf{Samuel Blazquez}: Investigation (equal); Methodology (equal); Writing – review \& editing (equal). \textbf{Eduardo Sanz}: Investigation (equal); Methodology (equal); Writing – review \& editing (equal). \textbf{Carlos Vega}: Conceptualization (equal); Investigation (equal); Methodology (equal); Writing – review \& editing (equal). \textbf{Felipe J. Blas}: Conceptualization (lead); Investigation (lead); Methodology (equal); Supervision (equal); Writing – original draft (equal); Writing – review \& editing (equal).

\section*{Data availability}

The data that support the findings of this study are available within the article.

\bibliography{bibfjblas}

\end{document}